\documentclass[12pt,preprint]{aastex}

\usepackage{color}

\slugcomment{Accepted for publication in The Astronomical Journal}

\shorttitle{Star Formation Across W3}
\shortauthors{Rom\'an-Z\'u\~niga et al.}

\begin{document}


\title{Star Formation Across the W3 Complex}


\author{Carlos G. Rom\'an-Z\'u\~niga\altaffilmark{1}, Jason E. Ybarra\altaffilmark{1}, Guillermo D. Megias\altaffilmark{2}, Mauricio Tapia\altaffilmark{1}, Elizabeth A. Lada\altaffilmark{3} and Jo\~ao F. Alves\altaffilmark{4}}

\altaffiltext{1}{Instituto de Astronom\'ia, Universidad Nacional Aut\'onoma de M\'exico, Unidad Acad\'emica en Ensenada, Km 103 Carr. Tijuana--Ensenada, Ensenada 22860, Mexico}
\altaffiltext{2}{Facultad de F\'isica. Universidad de Sevilla. Dpto. F\'isica At\'omica, Molecular y Nuclear, Sevilla 41080, Spain}
\altaffiltext{3}{Astronomy Department, University of Florida, 211 Bryant
Space Sciences Center, FL 32611, USA} 
\altaffiltext{4}{Institute of Astronomy, University of Vienna, T\"urkenschanzstr. 17, 1180 Vienna, Austria}


\begin{abstract}
We present a multi-wavelength analysis of the history of star formation in
the W3 complex. Using deep, near-infrared ground-based images, combined with images obtained with Spitzer and Chandra observatories, we identified and classified young embedded sources. We identified the principal clusters in the complex, and determined their structure and extension. We constructed extinction-limited samples for five principal clusters, and constructed K-band luminosity functions (KLF) that we compare with those of artificial clusters with varying ages. This analysis provided mean ages and possible age spreads for the clusters. We found that IC 1795, the centermost cluster of the complex, still hosts a large fraction of young sources with circumstellar disks. This indicates that star formation was active in IC 1795 as recently as 2 Myr ago, simultaneous  to the star forming activity in the flanking embedded clusters, W3-Main and W3(OH). A comparison with carbon monoxide emission maps indicates strong velocity gradients in the gas clumps hosting W3-Main and W3(OH) and show small receding clumps of gas at IC 1795, suggestive of rapid gas removal (faster than the T Tauri timescale) in the cluster forming regions. We discuss one possible scenario for the progression of cluster formation in the W3 complex. We propose that early processes of gas collapse in the main structure of the complex could have defined the progression of cluster formation across the complex with relatively small age differences from one group to another. However, triggering effects could act as catalysts for enhanced efficiency of formation at a local level, in agreement with previous studies.

\end{abstract}


\keywords{stars: formation --- stars: luminosity function --- stars: clusters}



\section{Introduction \label{s:intro}}

\par In the current picture of star formation, Giant Molecular Clouds (GMC) are
highly inefficient factories, in which only a small fraction of the available gas is converted into stars. Moreover, star forming regions in GMC complexes are highly heterogeneous. There is a considerable diversity among the stellar aggregations they produce. Differences are evident even at the level of their basic properties, e.g. sizes, numbers and density structures.

\par Aggregations of young stars in molecular clouds are classified as Embedded Star Clusters. These clusters are born as bound systems of gas and stars. Once gas is removed, embedded clusters may or may not stay bound. They either survive (e.g. open clusters), or dissolve to become part of the field population. For a majority of clusters in the galaxy, the latter is the common outcome \citep[referred as the ``infant mortality effect",][]{Lada:2003aa}. However, recent discussions consider that cluster morphology is more heterogeneous \citep{Bressert:2010fk,de-Grijs:2011rt}. Numerical studies suggest that small groups of stars may be abundant \citep{Adams:2001aa} and they may either merge to form clusters or disrupt rapidly into the field. According to studies like that of \citet{Kruijssen:2012ve}, star clusters may assemble from sub-clusters that merge into larger entities after the gas from which they formed is both disrupted by stellar feedback and torn apart by tidal shocks from the surrounding cloud (the so called ``cruel cradle" effect). 

\par Unfortunately, the dynamical picture of cluster evolution provided by numerical studies is difficult to directly compare with observations. On one hand, the initial conditions of cluster formation, a key requirement for realistic simulations, are still under debate. On the other hand, young cluster evolution from observations, would require of additional information on the kinematics of stars (e.g. radial velocities), and spectroscopic age estimations of individual sources to infer the progression of formation. Such studies are only beginning to arise from surveys like APOGEE\footnote{Sloan Digital Sky Survey, Apache Point Galactic Evolution Experiment} \citep[e.g.][]{cottaar:2014aa}, and are limited to nearby ($d<1$ kpc) regions. Meanwhile, it is possible to partially reconstruct the history of star formation in a region from photometric information, which can provide evolutive classification and the spatial distribution of young sources, and from molecular gas emission maps, which can provide gas distribution and kinematics.

\par Surveys of GMCs show that embedded star clusters are rarely (if ever) born in isolation. Most embedded clusters form as part of ``families" related to a particular complex, defining together a history of star formation from various levels of interaction. For instance, massive stars of one cluster may have influence on the efficiency of star formation of a neighboring cluster in the same cloud. Considering all this, it should be clear that star formation is highly dynamic; the embedded populations we observe are snapshots of a rather convoluted process of evolution and interaction, that changes significantly along the star forming history of a complex.

\begin{figure*}
\centering
\includegraphics[width=6.0in]{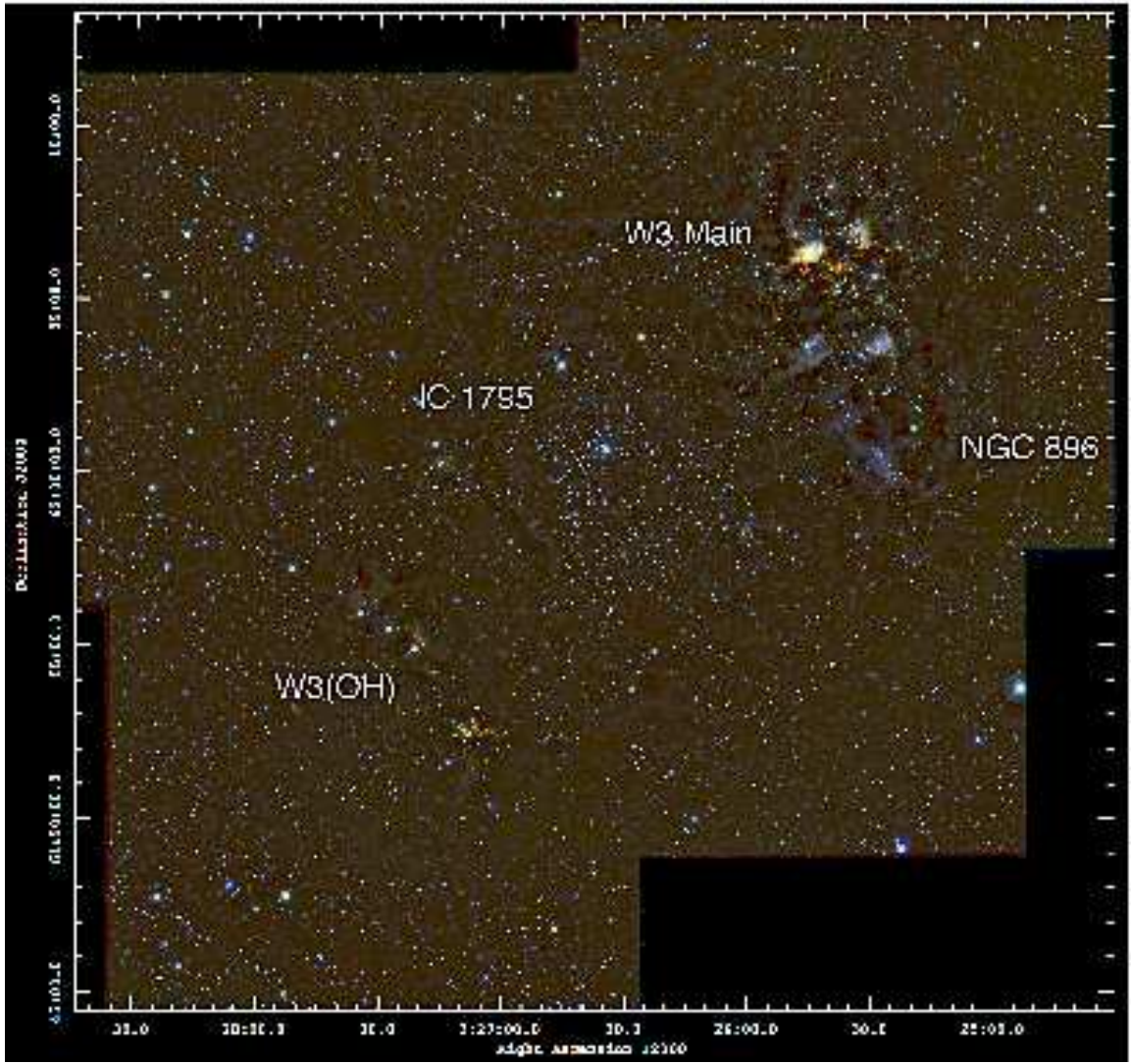}
\caption{A false color mosaic made from J (blue), H (green) and K (red) images from our OMEGA 2000 observations. . \label{fig:JHKmosaic}}
\end{figure*}

\par In order to investigate the progression of cluster formation under the influence of the local environment, we considered W3, a prominent cluster forming region in the ``Heart and Soul" molecular complex. W3 is considered a clear example of sequential formation, where cluster formation was induced by the expansion of the giant HII region W4 \citep{Lada:1978aa,Thronson:1985aa}. One of our goals is to trace the formation of distinct embedded clusters in this region and to attempt to reconstruct the star forming history of the cloud. The W3 Complex is located at a distance of 2.04$\pm$0.107 kpc \footnote{based on Very Long Baseline Interferometry (VLBI) parallaxes to H$_2$O masers in W3(OH) by \citet{Hachisuka:2006dz}}, and has hosted at least three major episodes of recent cluster formation. According to spectroscopic studies, IC 1795 formed first, about 3-5 Myr \citep{Oey:2005ly} ago, followed by the W3 ``main" cluster located to its West edge, and the W3(OH) cluster group to the East, both with ages of 2-3 Myr \citep{Bik:2012aa,Navarete:2011ys}. It has been suggested that IC 1795 triggered the other two episodes in a hierarchical progression \citep{Oey:2005ly}. The Chandra study by \cite{Feigelson:2008vn} suggested that the clusters in W3 extend widely and are highly structured, with sources located at relatively large distances from the dynamical centers, including a relatively isolated O star that might have escaped from the main cluster. 

\par In two recent studies, \citet{Rivera-Ingraham:2011yq,Rivera-Ingraham:2013fj} made use of mid-infrared photometry and far-IR emission mapping from the Spitzer and Herschel\footnote{Herschel is an ESA space observatory with science instruments provided by European-led Principal Investigator consortia and with important participation from NASA.} space observatories, that comprise the entire W3 region. They were able to catalog hundreds of young stellar sources (YSOs) across the complex and determine their spatial distribution. They compiled important evidence that projected distances among YSOs are consistent with cluster forming clump scales, favoring the cluster forming mode. They also concluded that small aggregations and distributed populations, account for a significant fraction of the recent stellar production in the region. They suggest relative large age spreads from central to external regions in W3 and proposed a ``convergent constructive feedback" scenario, where the gas flows from massive star formation clumps and collects into new dense regions, favoring a progression of formation and the age spread. 

\par In this study we emphasize the relative importance of different young stellar groupings in and around IC 1795, W3-Main and W3(OH), augmenting the level of detail achieved in previous studies. We present a new set of deep, high resolution near-IR photometry that we combine with other available datasets, allowing us to increase the number of young star candidates in the complex. We use a K-band luminosity function analysis to investigate the presence of an age spread, and to help reconstruct the history of star formation in the cloud. Finally, we investigate gas kinematics near the clusters, providing additional information on the gas-star interaction during the early evolution of the complex. 

\section{Observations and Data Reduction \label{s:observations}}

\subsection{Near-infrared imaging \label{s:observations:ss:nir}}

\par We obtained near-infrared images of four fields in W3, covering a large area ($0.58\deg\times$$0.46\deg$). Images were obtained with the Omega 2000 camera at the 3.5m telescope in the Calar Alto Observatory of the Centro Astron\'omico Hispano Alem\'an (CAHA) in Almer\'ia, Spain. Omega 2000 provides a $15^\prime \times 15^\prime$ field of view (FOV). Observations were made in $J$, $H$ and $K$ (1.1, 1.6 and 2.2 $\mu$m, respectively).  A list of all fields observed is listed Table \ref{tab:obs}, which gives the field identification, the center of field positions, observation date, filter, seeing (estimated from sigma-clipped average full width half-maximum (FWHM) of the stars in each field), and the peak values for the brightness distribution, which is a good estimate of the sensitivity limits achieved. The brightness distribution peaks are in all cases at or above $J=20.5$ mag, $H=19.75$ mag and $K=19.25$ mag.

\subsubsection{Image reduction \label{s:observations:ss:nir:sss:reduction}}

\par The Omega 2000 images were reduced with modified versions of the FLAMINGOS near-infrared reduction and photometry/astrometry pipelines, which are built in the standard \texttt{IRAF} Command Language environment. One pipeline \citep[see][]{Roman-Zuniga:2006aa} processes all raw frames by subtracting darks and dividing by flat fields, improving signal to noise ratios by means of a two pass sky subtraction method, and combining reduced frames with an optimized centroid offset calculation. We used dark frames and dome flats obtained within 48 hours of each observation. The final combined product images were then analyzed with a second pipeline, \citep[see][]{Levine:2006ab}, which identifies all possible sources from a given field using the \texttt{SExtractor} algorithm \citep{Bertin:1996aa}. We improved the \texttt{SExtractor} detection efficiency in the near-IR images by using a Gaussian convolution filter and maximum deblending \citep[see][]{Bertin:1996aa}. The pipeline then performs \texttt{Daophot} PSF photometry \citep{Stetson:1987aa}, calibrates observed magnitudes to a zero point and finds accurate astrometric solutions. 

\par
\begin{figure*}
\centering
\includegraphics[width=6.0in]{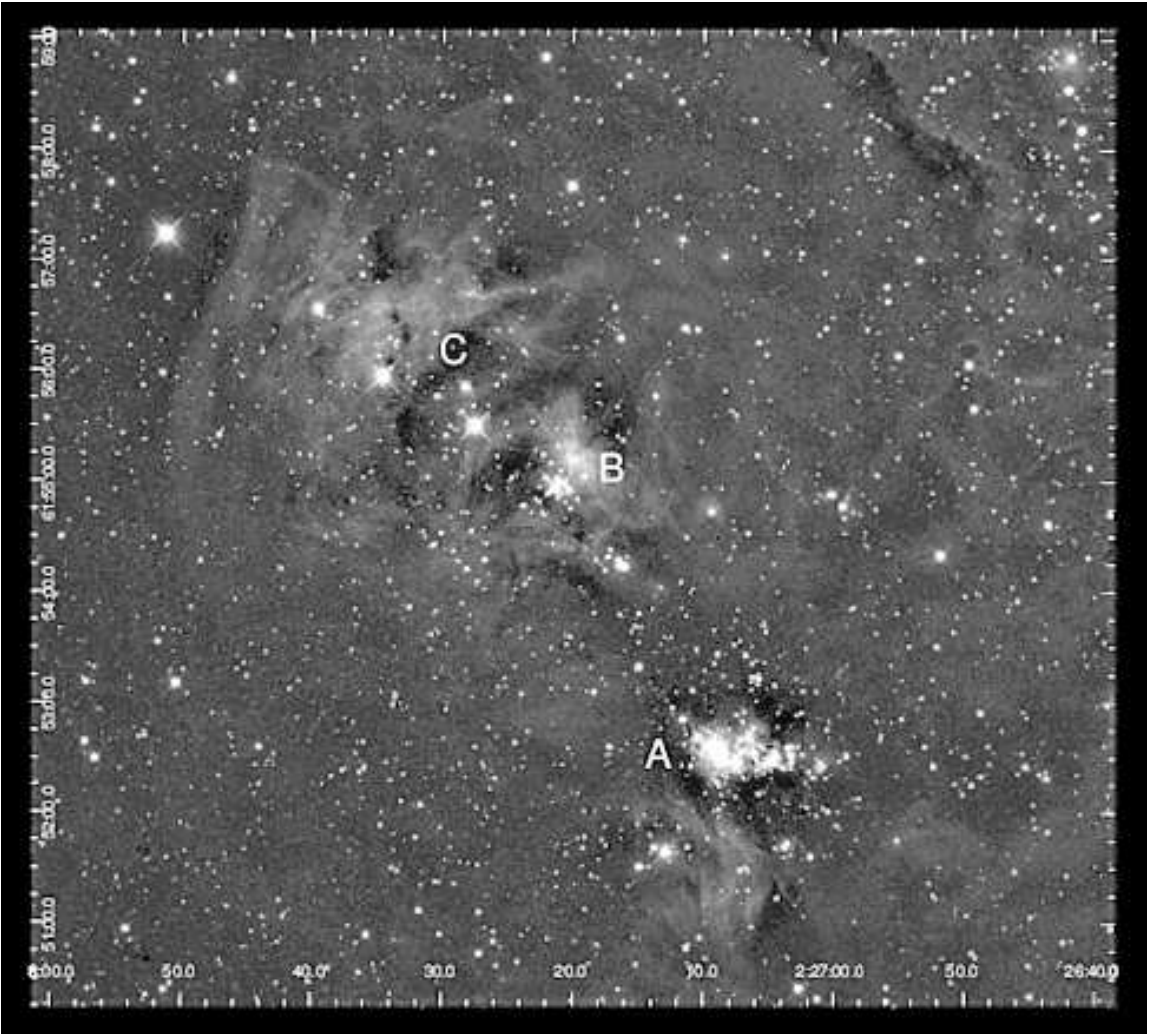}
\caption{RGB color mosaic of the W3(OH) region, from Spitzer IRAC 3.6 $\mu$m (red) and $K$ and $H$ OMEGA 2000 images (green and blue, respectively) showing possible sub-cluster structure in W3(OH). The W3(OH) cluster is labeled as ``A", while two possible sub-clusters are labeled ``B" and ``C". These last two labels are placed closer to B-type sources. \label{fig:W3OH}}
\end{figure*}

 After astrometric solutions were found, individual field images were combined into mosaics using \texttt{Montage} \footnote{http://montage.ipac.caltech.edu}. In Figure \ref{fig:JHKmosaic} we show a RGB panorama constructed from a combination of the $J,\ H\mathrm{\ and\ }K$ mosaics. The image represents a complete spatial coverage of the IC 1795, W3-Main and W3-OH clusters. In the near-infrared the region is transparent to most of the prominent regions of nebulosity and obscuration observable in optical images \citep[e.g. ][]{Ogura:1976aa}. The near-IR images at W3(OH) are particularly interesting because they reveal, with unprecedented detail, several small, embedded stellar groups that lie in a small ``chain" structure north and east of the W3(OH) cluster \citep[these groups have been previously identified; e.g.][]{Feigelson:2008vn,Navarete:2011ys}. In Figure \ref{fig:W3OH}  we show a close-up image of this region, where we now combine H and K images with the Spitzer IRAC 3.6 $\mu$m image to enhance illuminated nebulosity features and highlight the most reddened sources. We have labeled the W3(OH) cluster with the letter `A'. The two other  most conspicuous groups in the 'chain', both associated with B-type stars \citep{Navarete:2011ys} have been labeled with letters `B' and `C'. The former group is not as prominent as an over-density but as it can be seen in the mosaic, it is associated with thick nebulosity, so it is possibly more deeply embedded. There is one more bright source sitting at the center of a cavity in between groups B and C, which we suspect could be an additional sub-group.

\subsubsection{Catalog preparation \label{s:observations:ss:nir:sss:catalogs}}

\par The OMEGA 2000 photometry was calibrated relative to the Two-Micron All Sky Survey (2MASS), using catalogs retrieved from the All Sky Release Point Source databases. Final photometry catalogs were prepared with aid of \texttt{TOPCAT-STIL} \citep{Taylor:2005uq}. We combined individual photometry catalogs from each frame and filter and merged catalogs from all four fields into a master photometry list. In the overlapping areas, we selected the duplicates with the smaller total photometric error across the three bands in either field. Finally, we added 2MASS entries to replace failed measurements from saturated sources. Our final near-IR catalog contains a total of 72168 sources.

\subsection{Mid-infrared Spitzer imaging \label{s:observations:ss:mir}}

\par We performed a new reduction and photometric extraction of the W3 mosaics from project 1-127 available in the Spitzer Heritage Archive. Mosaics were constructed by processing both the short and long exposure (0.4 and 10.4 sec) Basic Calibrated Sets (BCD) from IRAC (all four bands) and the BCD set from MIPS (24 $\mu$m only) observations using the MOPEX package\footnote{http://irsa.ipac.caltech.edu/data/SPITZER/docs/\\dataanalysistools/tools/mopex/}. We used the IRAC short exposure datasets to improve extraction of sources down the centers of W3-Main and W3(OH) regions, which present large areas of saturation (increasing with wavelength) due to bright nebulosities in the long exposure sets. However, even at the short exposure images, the bright nebulosity  made it very difficult to obtain uniform quality from aperture photometry in the IRAC images, so we performed PSF photometry with \texttt{Daophot} on detection lists obtained with \texttt{SExtractor}. In this case, we improved the detection efficiency by using a ``Mexican hat" convolution filter and reduced deblending \citep[see][]{Bertin:1996aa} (compare to the near-IR, in section \ref{s:observations:ss:nir:sss:reduction}, above). Then, our pipeline selects a group of moderately bright and relatively isolated sources in order to make a PSF model for each \texttt{Daophot} run. In all cases we were able to fit the PSF model to a large majority of the detected sources. Final source detection lists were slightly cleaned with the aid of \texttt{PhotVis} \citep{Gutermuth:2004fr} to remove spurious detections.
\par The resultant photometry lists showed a net reduction of up to 12\% in the photometric scatter to the aperture photometry obtained with the APEX software. For the MIPS 24 $\mu$m mosaic we used the APEX aperture photometry pipeline, which was good enough for non-saturated regions of the field. In heavily saturated regions, particularly near the centers of W3-Main and W3(OH), missing pixels impeded the ability to make good measurements, and we could not improve these measurements with PSF photometry. Zero points for our Spitzer photometry catalogs were checked by direct comparison of our photometry lists with values from the GLIMPSE 360 Legacy Project catalog\footnote{only IRAC [3.6] and [4.5] bands are available as this is a warm mission extension of the original Galactic Legacy Infrared Midplane Survey Extraordinaire (GLIMPSE) project} and from the photometry tables of YSO candidates in \cite{Rivera-Ingraham:2011yq}. The agreement between both GLIMPSE and \cite{Rivera-Ingraham:2011yq} is excellent and zero point corrections were not required. The final photometry catalogs from the IRAC and MIPS 24 $\mu$m mosaics were then combined into a mid-IR catalog with 37267 sources, which we later merged with the near-IR catalog from our CAHA dataset to form a master source catalog.

\subsection{Other datasets \label{s:observations:ss:other}}

\par In this study we also make use of other datasets available in literature and public, web-based archives:
\begin{enumerate}
\item From the Chandra Source Catalog \citep[CSC; ][]{Evans:2010aa} the positions and X-ray photometry of a total of 611 sources in a box of 30$^\prime$ around the center of IC 1795\footnote{Observation ID numbers for W3 ACIS-I data are 446, 611, 7356, 5889, 5890, 5891, 6335 and 6348}.
\item The $^{12}$CO(2-1) and $^{13}$CO(2-1) maps of the W3 region from the study of \cite{Bieging:2011kq}. 
\item The 2.5-level (science grade)  SPIRE and PACS mosaics of the W3 region from the Herschel Space Telescope data archive. 
\item The Bolocam 1.1 mm mosaic of W3 from the the Canadian Galactic Plane Survey \citep[GPS; ][]{Aguirre:2011aa}. 
\end{enumerate}

\section{Data Analysis and Results \label{s:analysis}}

\subsection{Identification of Young Stellar Sources \label{s:analysis:ss:ysoid}}

\par Using our master infrared photometry catalog we identified Young Stellar Objects (YSOs) in the W3 complex. First, we identified Class I/0 and Class II sources from IRAC colors using the criteria applied by \cite{Ybarra:2013kh}, which are in turn based in the IRAC color criteria of \citet{Gutermuth:2008uc}. Then, using the Chandra CSC catalogs we identified Class III sources as X-ray sources with infrared counterparts and no excess associated to circumstellar material. Finally, we identified additional Class I and Class II sources as X-ray CSC sources with infrared excess, by applying criteria that combine near-infrared and IRAC colors \citep[see appendix of\ ][]{Gutermuth:2008uc}.

\par Figure \ref{fig:hardness} shows a $H-K$ vs. $K-[4.5]$ color-color diagram for all infrared sources that coincide in position with X-ray CSC sources. The colors of the symbols are indicative of the Hardness ratio, calculated from the hard and soft X-ray fluxes, $H$ and $S$, estimated from aperture photometry measurements, as $(H-S)/(H+S)$. It is clear that most of the sources have large hardness ratios, which indicates that in most cases X-ray sources are obscured by large amounts of dust which reduces the soft X-ray emission.

\begin{figure}
\centering
\includegraphics[width=5.5in]{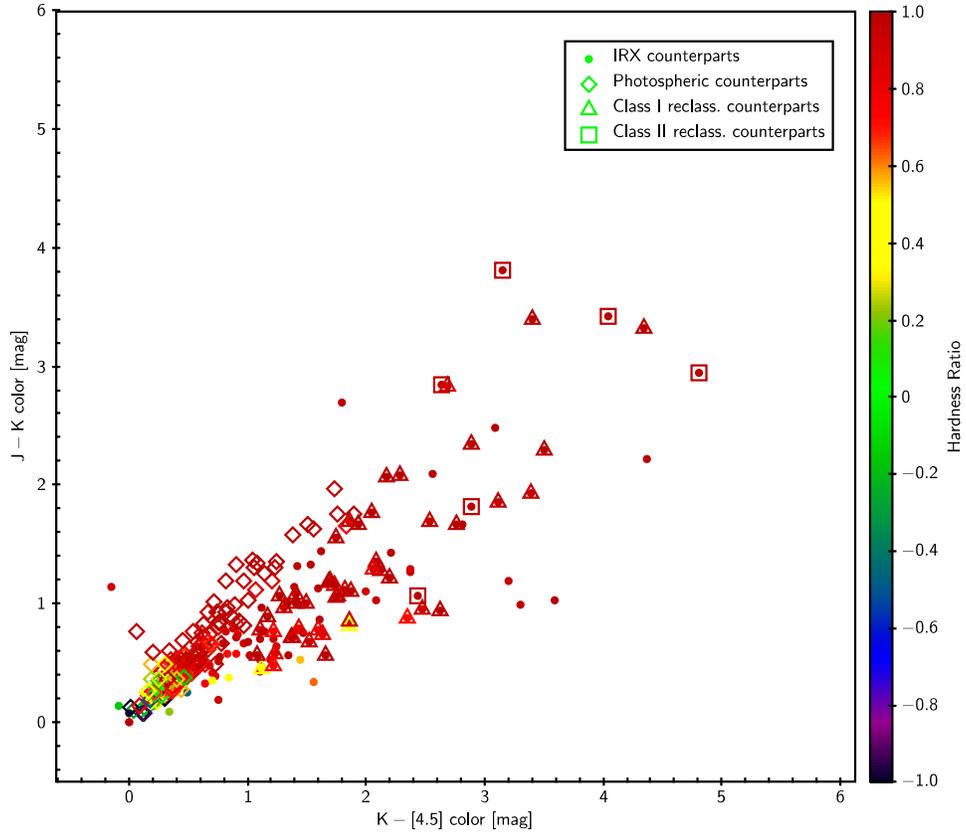}
\caption{$H-K$ vs. $K-[4.5]$ color-color diagram for CSC X-ray sources coinciding with an source in our infrared master catalog. Open diamond symbols are sources without an infrared excess indicative of a disk. Open square symbols indicate sources classified as Class I. Open triangle symbols indicate sources classified as Class II. The colors of the symbols are coded to the hardness-to-soft X-ray ratio, as indicated in the color bar.  \label{fig:hardness}}
\end{figure}

\par We identified a total of 1008 YSOs in W3, distributed as follows: 60 Class I/0 candidate sources, 780 Class II candidate sources and 167 Class III candidate sources. Our list contains a significantly larger number of YSO candidates compared to the list of \cite{Rivera-Ingraham:2011yq}. We suspect that the differences are mostly due to the addition of X-ray and near-infrared criteria which increased the number of candidates in each class. \citeauthor{Rivera-Ingraham:2011yq} only list candidates with detections in all IRAC bands, which complicates a direct assessment of completeness. 

\begin{figure}
\centering
\includegraphics[width=5.5in]{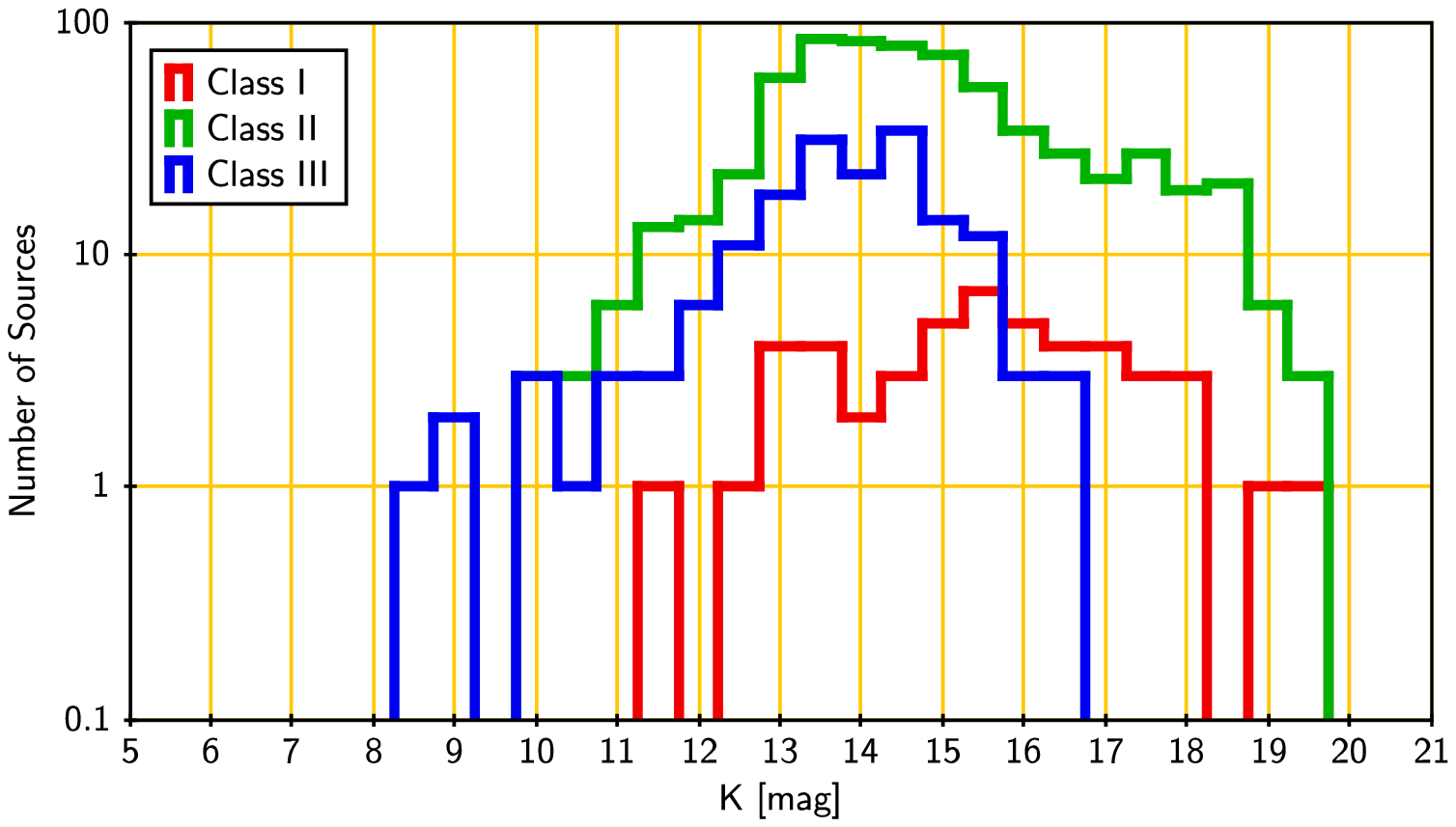}\\
\includegraphics[width=5.5in]{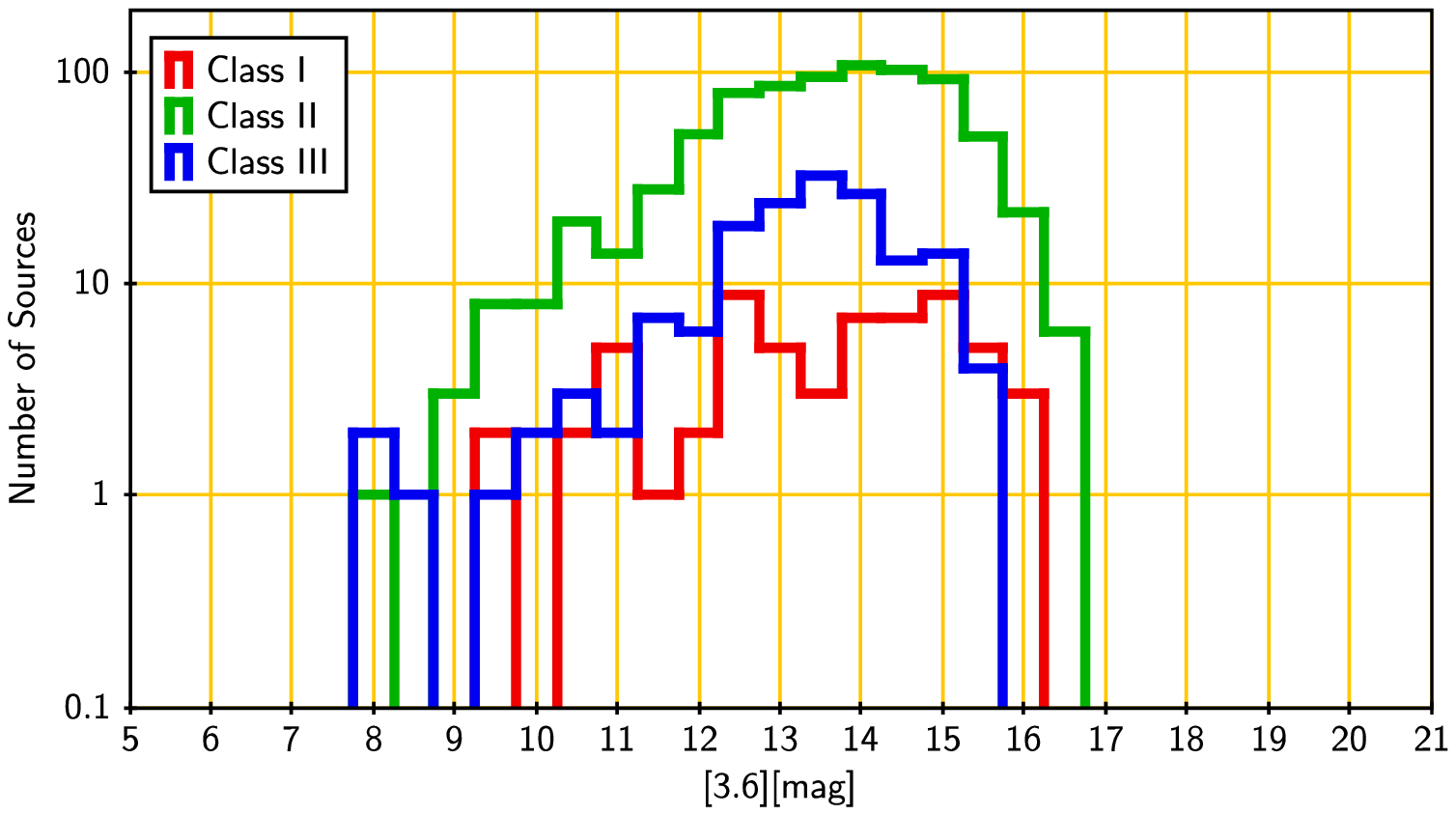}
\caption{Above: K band brightness distribution for young stellar source candidates identified in the W3 region. Below: [3.6] brightness distribution. \label{fig:ysobd}}
\end{figure}

In Figure \ref{fig:ysobd} we show the K and [3.6] band brightness distributions of the three identified YSO classes in our sample. In the case of the [3.6] band, the three distributions show a sharp drop at a similar limit of about 15.5 mag. In contrast, for the $K$ band, the distribution for Class III sources drops at about 15.0 mag, while the distributions for Class I and Class II sources drop near 18.0 mag. These histograms show that, due to the high and non-uniform extinction, the brightness distributions for young sources are likely incomplete above the sensitivity limits. However, our near-IR observations are deep enough to detect young sources across the entire region. For this reason, analysis of the luminosity functions discussed in Section \ref{s:analysis:ss:klf} were performed using extinction-limited samples. 

\subsection{Spatial Distribution of Young Stellar Sources \label{s:analysis:ss:ysodist}}

\par
\begin{figure*}
\centering
\includegraphics[width=6.0in]{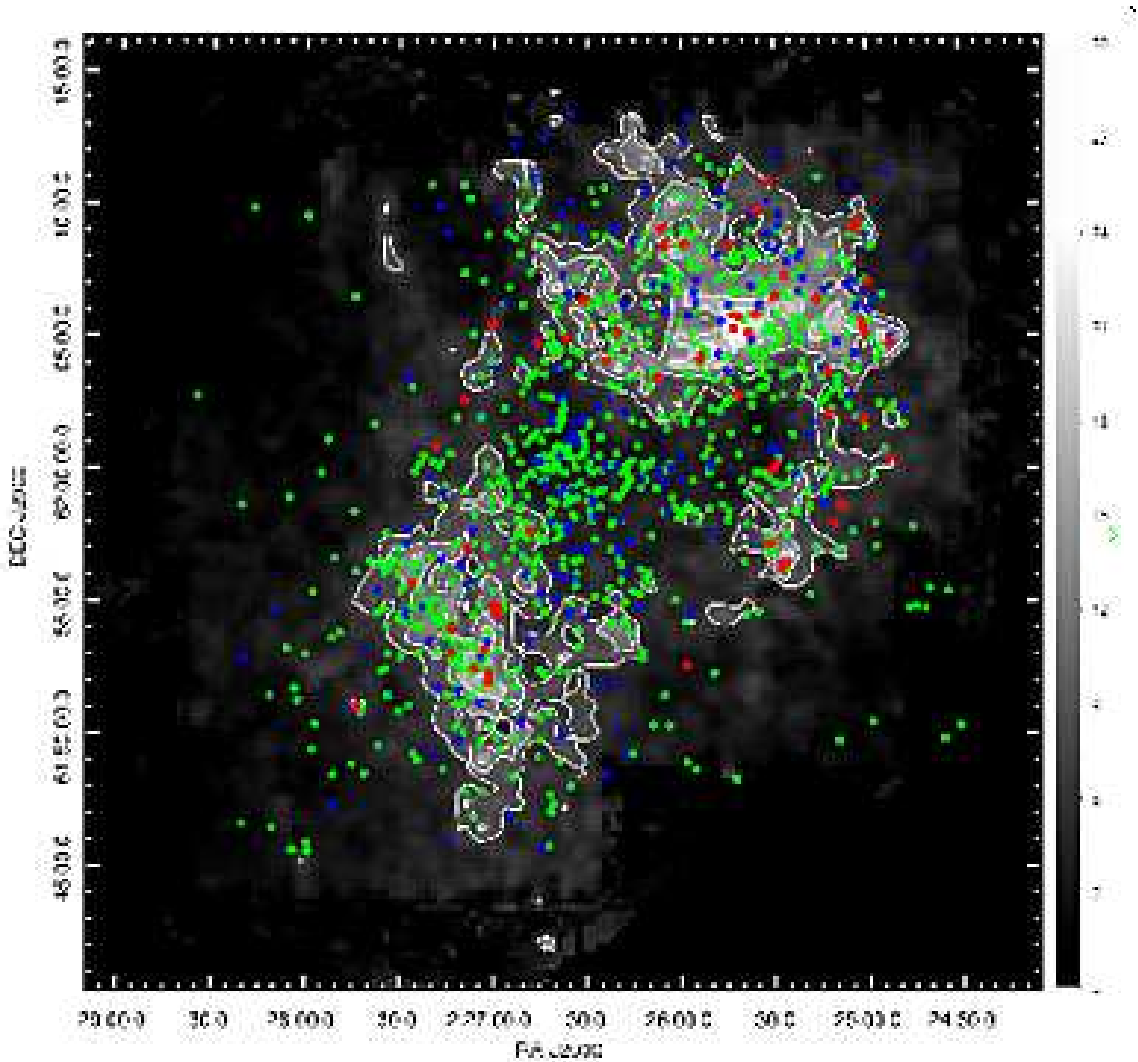}
\caption{The plot shows the positions of infrared young stellar sources in the entire W3 region. Red symbols indicate sources classified as Class I. Green symbols indicate Class II. Blue symbols indicate Class III sources. The background image and contours correspond to a visual extinction ($A_V$, in units of magnitudes) map of the region constructed with the NICEST method. The map is Nyquist sampled over a 18$\arcsec$ pixel grid, smoothed with a Gaussian filter with a 36$\arcsec$ FWHM. The contours indicate extinction at $A_V=$7.0, 10.0, 15.0, 20.0 and 25.0 in units of magnitudes. \label{fig:ysopos}}
\end{figure*}

 In Figure \ref{fig:ysopos} we illustrate the spatial distribution of YSO candidates in the W3 region, overlaid on an dust extinction map constructed with the optimized near-infrared excess method NICEST method of \citet{Lombardi:2009aa}. The YSO sources appear to be very well constrained to the three well known areas of the complex (namely IC 1795, W3-Main and W3(OH)). 
\par Class I sources are mostly confined to the areas of high extinction: following the prescription of \citet{Lada:2013aa}, we found that over 80 percent of the Class I sources are located in high extinction regions ($A_V>7.0$ mag), and the surface density of Class I sources is linearly correlated with $A_V$ within $3<A_V<12$ mag. This is in very good agreement with other cluster forming regions like Orion A  \citep{Lada:2013aa} and the Rosette Molecular Cloud \citep{Ybarra:2013kh}. 

\par Class II sources clearly permeate the entire complex, while Class III sources appear to have a slightly more centrally condensed distribution in the IC 1795 region but are also present in the western edge of W3-Main and surrounding the areas of low extinction to the East of W3(OH). 

\par Two aspects are important to notice. First, cluster regions are not separated clearly in the layout of the complex, which suggest at least some partial overlap along the line of sight perpendicular to the field. Second, the spatial distribution of young sources suggest that clusters are not single, centrally condensed systems. Instead, they appear to have sub-structure.

\par In the following we a) make a consistent separation of the individual cluster populations and b) provide evidence of the presence of sub-structure in the cluster population by using a surface density analysis. We make the following two definitions:

 \begin{enumerate}
\item We define a \textit{Principal Cluster} as a surface density structure that can be associated to a single Gaussian peak with a FWHM of at least 1.0 pc (see $\S$\ref{s:analysis:ss:ysodist:sss:GMM}, below). 

\item We define a \textit{sub-structure} as a significant surface density peak, smaller than a principal cluster. We identified as significant, all those peaks with a surface density above 1.0 stars per sq. arcmin (see $\S$\ref{s:analysis:ss:ysodist:sss:gather}, below).
\end{enumerate}

\subsubsection{Identification of Principal Clusters \label{s:analysis:ss:ysodist:sss:GMM}}

\par In order to delimit individual clusters, we employed a Gaussian Mixture Model (GMM) analysis via the \texttt{mclust} package on the \texttt{R} code \citep{Banfield:1993xy,Fraley:2002qf,Raftery:2012nr}. The package provides a model-based clustering analysis of any population, based on covariance parametrization and an Expectation-Minimization (EM) algorithm. It can also divide a population in an optimized number of clusters by means of the Bayesian Information Criterion (BIC). The BIC is the value of the maximized log-likelihood with a penalty on the number of model parameters, and allows comparison of models with different parameter sets as well as solutions with different numbers of clusters.

\begin{figure}
\centering
\includegraphics[width=6.2in]{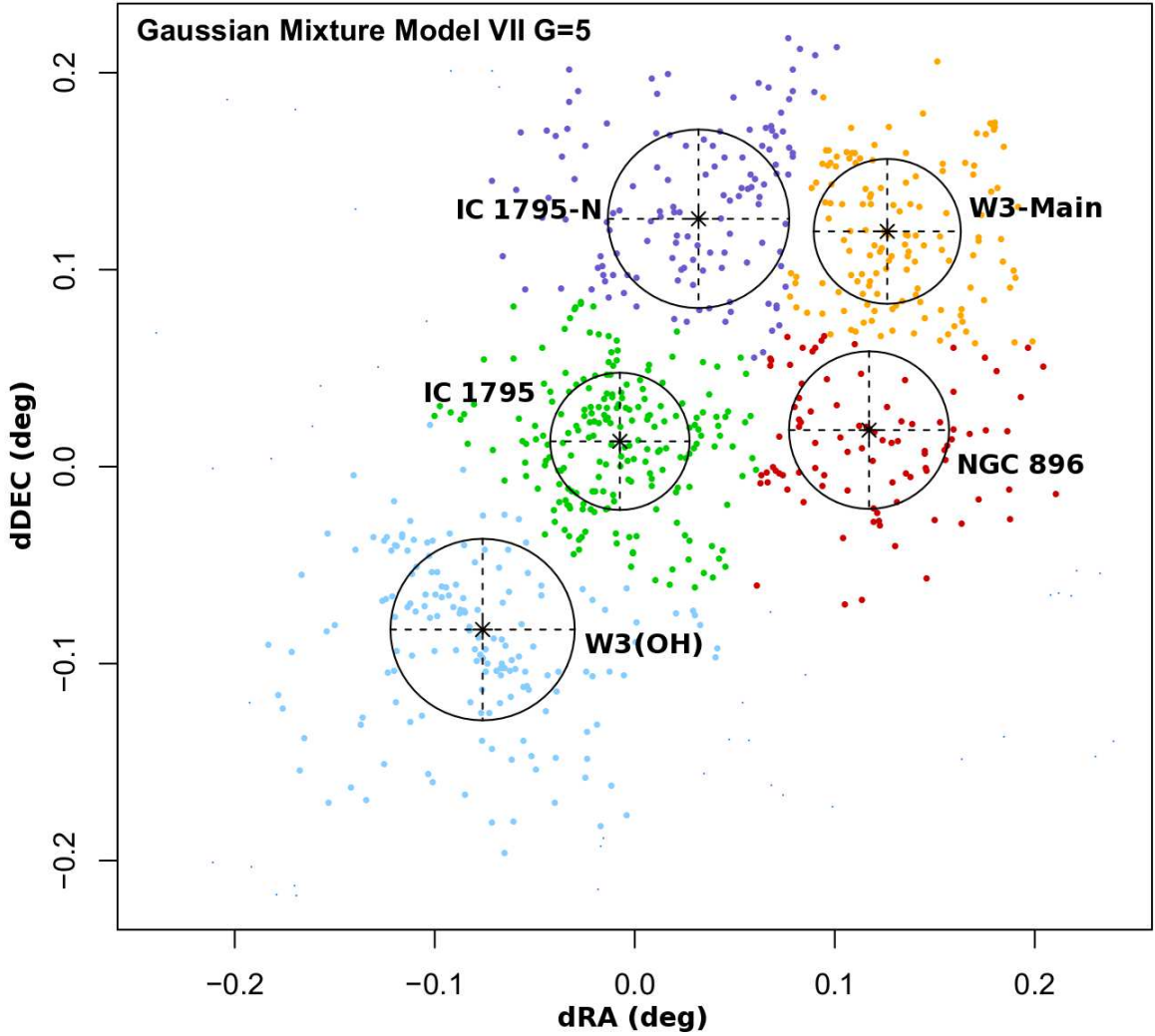}
\caption{Identification of principal clusters in the W3 region using a
Gaussian Model Mixture (GMM). This five cluster model, applied to projected
distribution of the population of Class II YSO candidates in the region
provides the optimal mixture based on a Bayesian Information Criterion (BIC). The circles are placed at the cluster centers determined by the GMM analysis and have radii equal to the Gaussian $\sigma$. \label{fig:GMM}}
\end{figure}

\par We chose the Class II sample as it is the most complete and it is distributed more or less uniformly across the complex. RA,DEC positions were fed into the code as a sample of bivariate observations. We chose to use circular Gaussians to minimize the number of parameters required for the estimation. Given the complicated layout of the clusters in W3, we opted to run the GMM analysis including noise modeling and rejection of outliers\footnote{ In our case, we chose to use a first order estimation of noise in the spatial distribution of the sources on the plane based on nearest-neighbors.  This way the \texttt{mclust} package applies the hierarchical clustering algorithm to a denoised dataset, and finally,rejects from the final solution, those sources that do not belong any cluster in the final solution \citep{Raftery:2012nr}.}. For our dataset, a model of five clusters provided the best fit for the sample. This is shown in Figure \ref{fig:GMM}. While this finite mixture model shows less detail than --for instance-- the nearest neighbor analysis, we see that it succeeds at identifying the three well known clusters in the region (W3-Main, W3(OH), IC 1795). It also adds two more  principal clusters, one coinciding with NGC 896, the compact HII region located at the Western edge of the complex, and one more located at the Eastern edge. We labelled this group IC 1795-N. In Table \ref{tab:principal_clusters} we list the result of the GMM analysis, with the centers and radii of the five principal clusters identified (columns 1 to 4).

\par We emphasize that the purpose of this exercise is to define the principal clusters as consistent samples to study the history of star formation in the complex. The GMM analysis provides such consistent identification,
which may differ with respect to other identification methods \citep[e.g.\ ][]{Rivera-Ingraham:2011yq}. 

\subsubsection{Cluster Substructure in W3. Nearest Neighbors. \label{s:analysis:ss:ysodist:sss:gather}}

\par Estimating the surface density of young stellar sources is often done by applying a variant of the nearest neighbor method \citep[formalized for star cluster detection by\ ][]{Casertano:1985uq}. This method has been applied successfully in different studies \citep[e.g.\ ][]{Gutermuth:2010aa,Roman-Zuniga:2008aa}. However, two very important issues regarding this method must be pointed out. First, when calculating distances to the $n$th neighbor in a distribution of points on a plane, there is a minimum value $n=6$ to avoid biasing the detection toward small structures formed purely by chance in a 2-dimensional distribution of points. Second, when mapping the surface density defined by such distances, the choice of an optimal smoothing kernel that would not dilute genuine substructures neither generate spurious ones, is not straightforward. These two issues were taken into consideration by \cite{Gladwin:1999id}, who developed two nearest neighbor mapping methods, \textit{gather} and \textit{scatter}, both optimized to correctly detect small groups (in our case, possible cluster substructures) in source position maps. The \textit{gather} algorithm is more adequate for determining hierarchical structure, while the \textit{scatter} method is more adequate for defining the size of the smallest group in a distribution of points on a map. 

\begin{figure*}
\centering
\includegraphics[width=6.0in]{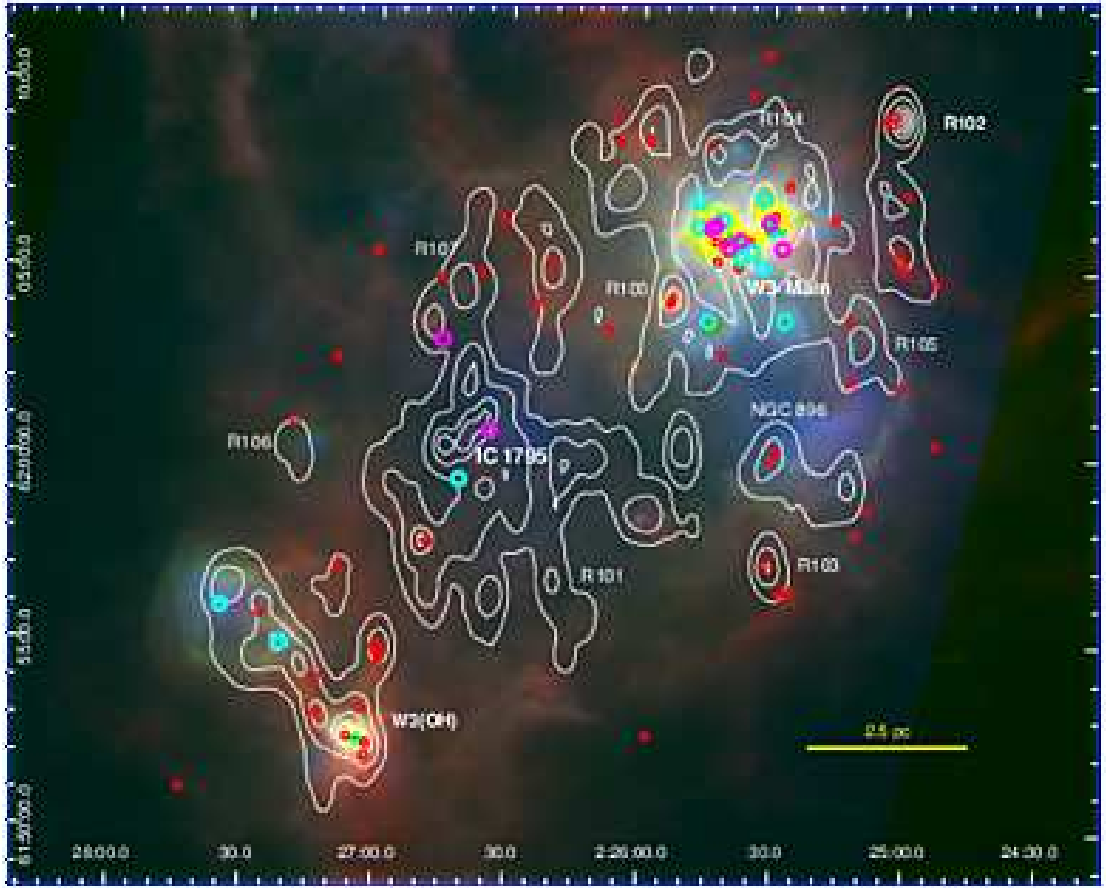}
\caption{RGB image constructed by combining map from Spitzer MIPS 24$\mu$m (blue), Herschel PACS 70 $\mu$m (green) and
Herschel SPIRE 250$\mu$m (red). Contours indicate surface density of Class II young stellar sources the W3 region, estimated with the \textit{Gather} method. Contour levels are set as 0.5,1.0,2.0,3.0,4.0,5.0 and 10.0 stars per square arcmin,. The positions of Class I sources are indicated with red circle symbols. The positions of O and B type stars are indicated with magenta and cyan color symbols, respectively. Areas in yellow/green are saturated pixels in the infrared images.  The labels correspond to previously identified clusters and sub-clusters in the W3 region; labels starting with letters ``RI" indicate those identified by \citet{Rivera-Ingraham:2011yq}. \label{fig:gather}}
\end{figure*}

\par The \textit{gather} algorithm was better suited for the task of detecting and separating sub-structures.  We applied \textit{gather} to the distribution of all Class II YSO candidates in the W3 region, which is the most complete of our samples. Densities were calculated with a square kernel of 36$\arcsec$ over a Nyquist sampled RA-DEC grid. In Figure \ref{fig:gather} we show the resultant  map, in the form of surface density contours overlaid on an infrared emission image that combines data from the Spitzer MIPS 24$\mu$m, 70$\mu$m from Herschel PACS and 250 $\mu$m from Herschel SPIRE. If we compare the density map to the spatial distribution of sources in Figure \ref{fig:ysopos} we see how \textit{gather} highlights the three main regions (IC 1795, W3-Main and W3(OH)) as the principal clusters, but also shows a number of peaks that clearly define significant sub-structure in the spatial distribution of young sources. 

\par The positions and equivalent sizes of sub-structure peaks in the \textit{gather} map are listed in Table \ref{tab:substructures}. Some of the sub-structure peaks in the map coincide in position with one or several Class I sources. Those peaks are likely related to the most recent episodes of formation in the complex. Notice how none of the peaks near the center, corresponding to IC 1795, are associated with Class I sources, possibly indicating an absence of current formation. We also show the positions of O and B type stars listed by  \citet{Ogura:1976aa}, \citet{Navarete:2011ys} and \citet{Bik:2012aa}. 

\par The largest concentration of massive stars (6 O-type stars, 7 B-type stars) is located at W3-Main, along with a noticeable concentration of protostars, associated with at least four YSO surface density sub-structures. In IC 1795 there are two O-type stars and one B-type star identified. The central 
O star in IC 1795 (No. 89 in the \citet{Ogura:1976aa} list) is associated with the highest surface density peak.  In the W3(OH) region, two sub-structures define the positions of groups B and C in W3(OH), and both are associated with B type stars. It is interesting that the main cluster (group A) in W3(OH) is not associated with any identified massive source.

\begin{deluxetable}{lcccc}
\tablecolumns{5}
\tablewidth{0pc}
\tablecaption{Sub-structures in W3 \label{tab:substructures}} 

\tablehead{
\colhead{No.} &
\multicolumn{2}{c}{Center Coords.} &
\colhead{Class I\tablenotemark{a}} &
\colhead{Notes}\\
\colhead{} &
\multicolumn{2}{c}{J2000} &
\colhead{} &
\colhead{} \\
}
\startdata

1  & 02:27:33.29 & 61:56:24.2 &  $\dots$   & $\dots$	\\
2  & 02:27:21.43 & 61:54:57.4 &  $\dots$   &  W3(OH)-C  \\
3  & 02:27:16.37 & 61:54:16.0 &  $\dots$   &  W3(OH)-B  \\
4  & 02:27:12.43 & 61:53:08.8 &     1 &  $\dots$  \\
5  & 02:27:03.46 & 61:52:33.3 &     3 &  W3(OH)-A  \\
6  & 02:27:18.20 & 62:00:04.1&     1 &  R106 \\
7  & 02:27:08.58 & 61:56:30.6 &     1 &  $\dots$ \\
8  & 02:26:58.74 & 61:54:53.8 &     4 &  $\dots$ \\
9  & 02:26:49.52 & 61:57:46.0 &     1 & $\dots$  \\
10 & 02:26:48.17 & 62:03:28.3 &  $\dots$   & $\dots$  \\
11 & 02:26:40.22 & 62:00:18.4 &  $\dots$   &  IC 1795 center \\
12 & 02:26:41.14 & 62:04:33.6 &  $\dots$   &  R107 \\
13 & 02:26:38.87 & 62:02:01.3 &  $\dots$   &  $\dots$ \\
14 & 02:26:38.87 & 62:00:42.2 &  $\dots$   &  $\dots$ \\
15 & 02:26:35.50 & 61:59:15.1 &  $\dots$   &  $\dots$ \\ 
16 & 02:26:33.25 & 61:56:25.0 &  $\dots$   &  $\dots$ \\  
17 & 02:26:20.01 & 62:05:01.3 &    1  & $\dots$  \\
18 & 02:26:16.96 & 61:59:48.7 &  $\dots$   &  $\dots$ \\ 
19 & 02:26:19.22 & 61:56:42.8 &  $\dots$   & R101 \\
20 & 02:26:03.36 & 62:07:13.7 &  $\dots$   &  $\dots$ \\ 
21 & 02:25:59.28 & 61:58:37.3 &  $\dots$   &$\dots$	\\
22 & 02:25:57.97 & 62:08:50.6 &    1  & $\dots$  \\
23 & 02:25:52.42 & 62:04:21.4 &    2  & R100  \\
24 & 02:25:51.10 & 62:00:20.0 &  $\dots$  &  $\dots$  \\ 
25 & 02:25:47.76 & 62:10:35.2 &  $\dots$     & R104 \\
26 & 02:25:42.74 & 62:08:10.7 &  $\dots$  &  $\dots$  \\
27 & 02:25:43.36 & 62:06:18.0 & $\dots$   &  W3-Main center  \\
28 & 02:25:32.91 & 62:06:49.4 & $\dots$   &  $\dots$  \\
29 & 02:25:32.98 & 62:04:32.9 &    2  & $\dots$ \\
30 & 02:25:30.88 & 62:00:05.7 &    2  &  NGC 896 center \\
31 & 02:25:30.41 & 61:57:11.6 &    3  & R103 \\
32 & 02:25:21.90 & 62:07:06.8 &  $\dots$  &  $\dots$  \\ 
33 & 02:25:10.23 & 62:03:11.0 &  $\dots$     & R105 \\
34 & 02:25:13.21 & 61:59:21.6 &  $\dots$  &  $\dots$  \\
35 & 02:25:00.93 & 62:09:08.7 &    3  & R102 \\
36 & 02:25:03.58 & 62:06:52.4 &    1  & $\dots$  \\
37 & 02:25:00.26 & 62:05:37.0 &    3  &  $\dots$  \\

\enddata
\tablenotetext{a}{Number of Class I candidates within \texttt{gather} surface density peak contour}
\end{deluxetable}

\par In Figure \ref{fig:gather} we also placed labels (``RI") next to regions coinciding with ``sub-clusters" identified by \citeauthor{Rivera-Ingraham:2011yq}. It is worth noticing how the \textit{gather} map highlights the groups $A,\ B,\ C$ that were identified by eye in Figure \ref{fig:W3OH}.

\subsubsection{YSO Ratio Maps \label{s:analysis:ss:ysodist:sss:ratiomaps}}

\par
\begin{figure*}
\centering
\includegraphics[width=4.2in]{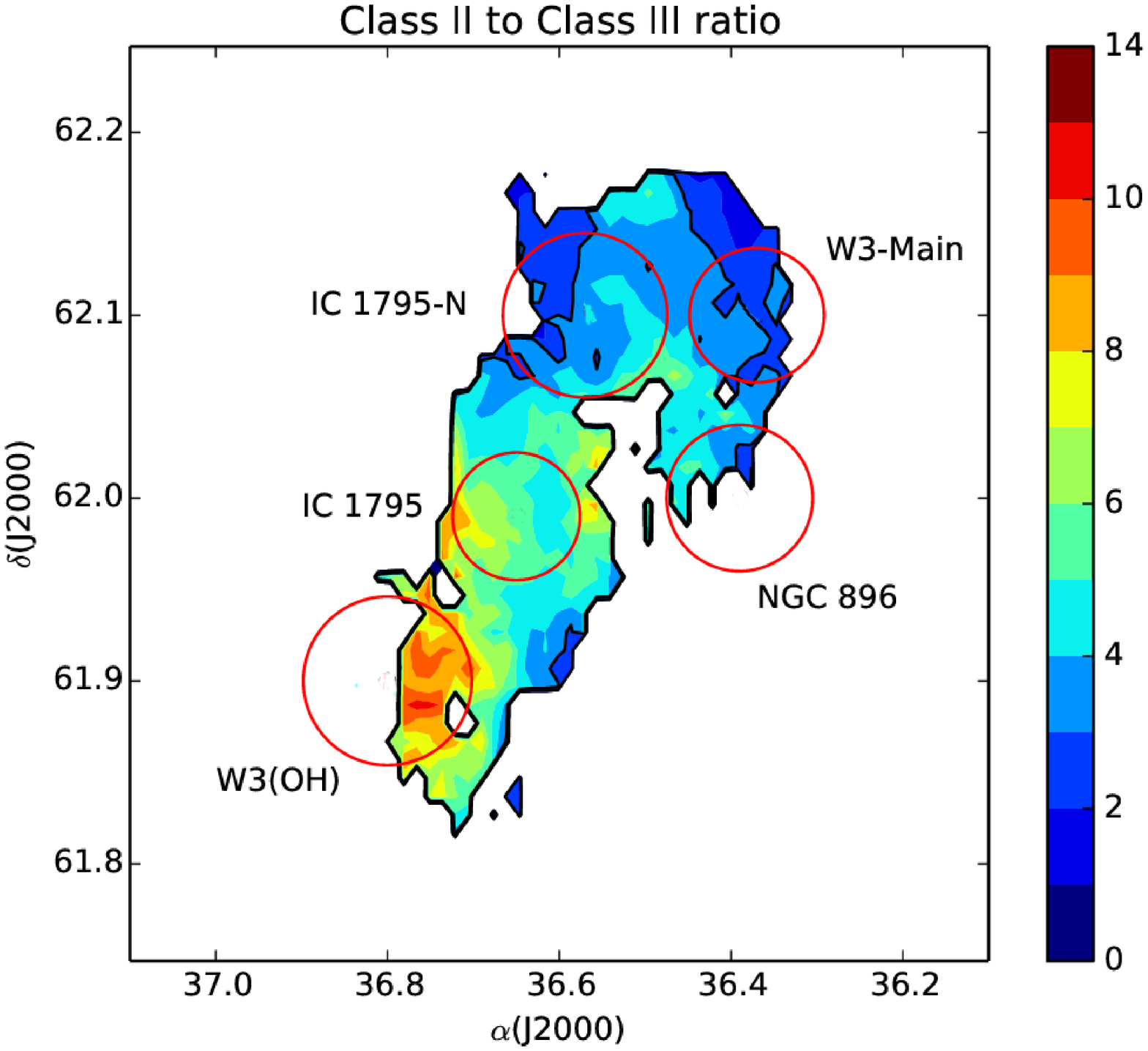}\\
\includegraphics[width=4.2in]{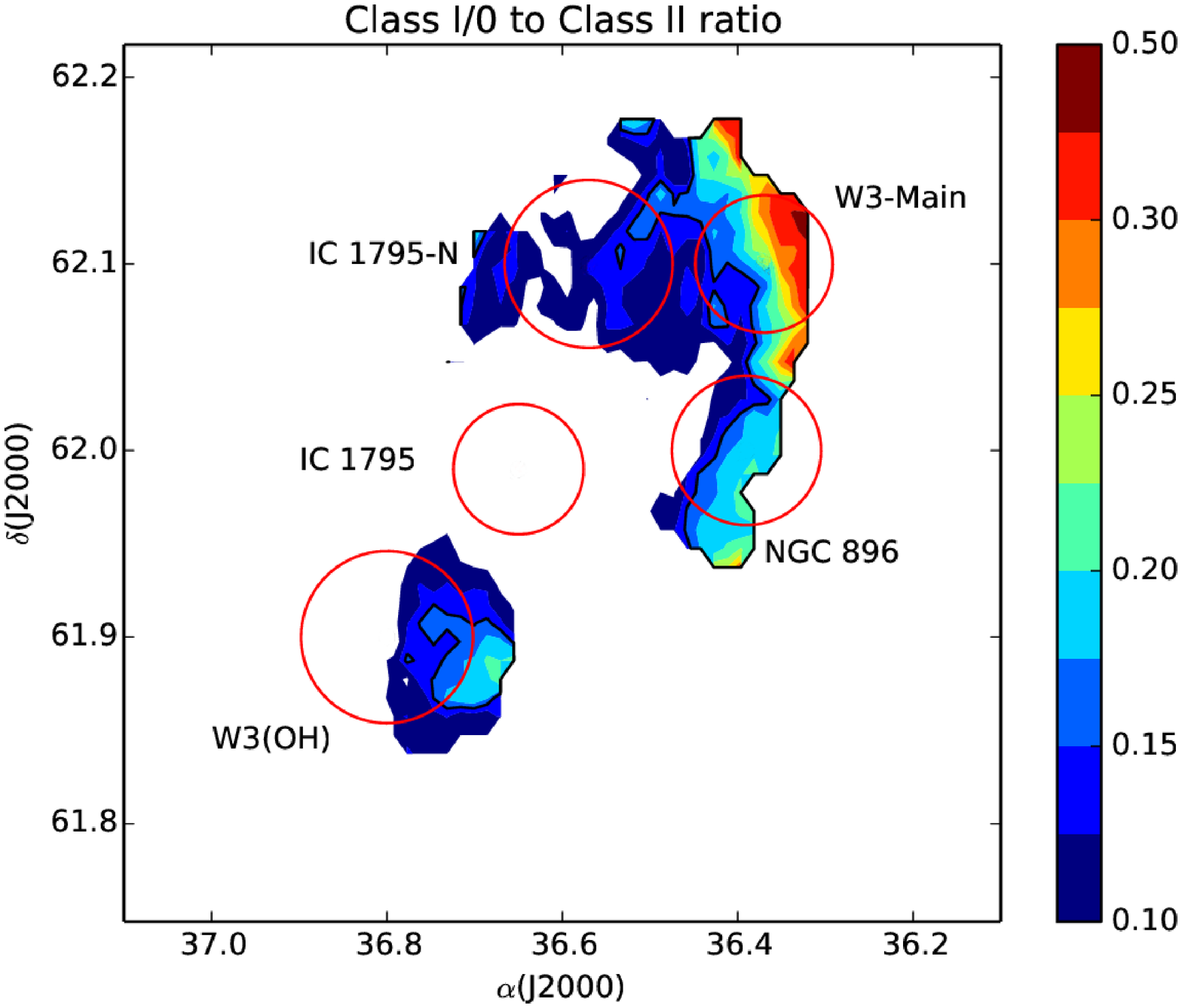}
 \caption{\textit{Top}: Class II to Class III ratio map. The black contours indicate $R_{\mathrm{II:III}}$ = {1.0,3.0}. Red circles indicate the positions and radii
 of principal clusters as in Fig. \ref{fig:GMM}. \textit{Bottom}: Class I/O to Class II ratio map.
 \label{fig:ratiomaps}}
\end{figure*}

 It is possible to trace the evolution of a star-forming region through the construction of maps of YSO number ratios. Figure \ref{fig:ratiomaps} shows maps for the Class II to Class III number ratio, $R_{\mathrm{II:III}}$, and the Class I/0 to Class II number ratio, $R_{\mathrm{I:II}}$, both constructed using the method described in \citet{Ybarra:2013kh}.

\par For the $R_{\mathrm{II:III}}$ map we limited the source candidates to a de-reddened brightness $H$=13.5 mag. This brightness cut avoids biasing the number counts from incompleteness due to patchy extinction. We used the Bayesian estimator described in \citet{Ybarra:2013kh} to calculate the ratios within circular projected regions with radius 3$^\prime$. The map only shows ratios for regions with sufficient number counts such that $N_{1}+N_{2}+1 <  0.25(N_{1}+1)(N_{2}-1)$, where $N_{1}$ is the number count of the numerator, and $N_{2}$ is the number count of the denominator.

\par We found that the whole complex has $R_{\mathrm{II:III}} \ge 1.0$ which
suggests star formation was ubiquitous throughout the complex $\sim$ 2-3 Myr ago. We can also trace the most recent star formation by using the $R_{\mathrm{I:II}}$ ratio (Fig. \ref{fig:ratiomaps}, right). Regions with significant $R_{\mathrm{I:II}}$ are expected to have been actively forming stars within the last Myr. We found that the regions of high $R_{\mathrm{I:II}}$ appear to form an oblong ring around IC 1795. This mirrors the distribution of Class I/0 sources which also appear to surround IC 1795 (see Fig. \ref{fig:gather}). Interestingly, the high $R_{\mathrm{II:III}}$ in IC 1795 suggests that stars were forming there up to 1 Myr ago. Therefore the absence of Class I/0 sources in IC 1795 suggests that star formation was recently and possibly abruptly terminated in that region.

\par It is important to notice that values for the $R_{\mathrm{II:III}}$ ratio in the map can be high (typically larger than 5), particularly because the Class III sources are more spatially scattered. However, the $R_{\mathrm{II:III}}$ is not the same as the circumstellar disk fraction in the cluster, because we are only measuring against weak disk sources and not the whole population (see also section \ref{s:analysis:ss:klf}, below).

\subsection{K band Luminosity Function of Clusters in the W3 Complex \label{s:analysis:ss:klf}}

\par We obtained the K band luminosity function (KLF) for each of the five principal clusters identified with our GMM analysis using the following procedure: 

\par Each cluster sample is composed by all K-band sources detected within a ``3-sigma" radius defined as $R_{3\sigma}=(3.0/\sqrt{8\ln 2})r_{cl}$, where $r_{cl}$ is defined from the GMM analysis as the half-width at half-maximum of the best fit 2-D Gaussian. Defining samples within $R_{3\sigma}$ minimizes effects of cluster overlapping, which could mislead in the estimation of age and age spreads in cluster forming regions \citep[e.g. NGC 1980,][]{Alves:2012aa}.

\par We obtained an extinction limited sample, defined as follows: 1) we projected a $A_V=$20 mag extinction vector from a 1 Myr isochrone \citep{DAntona:1997aa} until it reached the sensitivity limit of the observations. The $A_V$-limited sample is defined by all sources enclosed within the area defined by the extinction vector, the zero-age main sequence and the reddened isochrone (see Figure \ref{fig:avlimitedsample}, left). 2) we separated the possible foreground population from sources with colors coinciding with the locus of the dwarf and giant sequences in the $J-H$ vs $H-K$ color-color diagram.This defines a foreground contribution. In Figure \ref{fig:avlimitedsample} (right panel) we show a color-magnitude diagram for W3-Main showing the $A_V$-limited sample selection.  

\par We estimated the fraction of sources with circumstellar disks within $R_{3\sigma}$, for each principal cluster, as the total number of Class I plus Class II sources in the extinction limited sample, down to a de-reddened brightness $H$=13.5 mag. (see also section \ref{s:analysis:ss:ysodist:sss:ratiomaps} above). The highest fractions of circumstellar disks, close to 30 percent, are in W3-Main, IC 1795 and W3(OH). Clusters NGC 896 and IC1795-N have smaller fractions, closer to 15 percent.

\par The total contribution from the field was estimated from a nearby control field with minimum extinction, observed in the same conditions and to the same depth as the cloud fields (see Appendix \ref{app:obs}). We made a K-band histogram from the control field catalog and scaled it to the same area of the cluster sample. Then we subtracted the foreground contribution from the control field KLF, and  applied artificial reddening to the remaining distribution down to the same level as the cloud sample. Artificial reddening is applied by constructing a normalized $A_V$ extinction distribution from the background stars in each cluster region using the NICER algorithm, and used it as a frequency distribution to convolve the control field K-band histogram, producing a final control field K band brightness distribution. The NICER algorithm provides estimates of extinction with average errors $\sigma_{A_V}<$0.1 mag below $A_V=$30 mag. These errors are way below the size of the K-band brightness bins used to construct the KLF. The final KLF of the cluster sample is obtained by subtracting the final $A_V$-convolved (reddened) control field distribution from the one constructed from the K-band limited sample.

\par The advantages of using a K-band limited sample to construct the KLF are: a) that all samples are cut at common maximum extinction; b) that the contribution of background extra-galactic sources is reduced to a minimum; c) that patchy extinction is less prone to bias the brightness distribution, specially at the most embedded regions --e.g. centers of W3-Main and W3(OH). Also, by convolving the raw control field KLF by the ``on" field extinction distribution, we are able to obtain a fair estimation of the contribution from field stars on different lines of sight \citep{Muench:2003aa}.

\subsubsection{Age Inference from Cluster Model Fits \label{s:analysis:ss:klf:sss:ages}}

\par The KLF for each principal cluster sample was then compared to \textit{artificial} K-band brightness distributions of embedded cluster populations of varying ages, constructed with the model interpolation code of \citet{Muench:2000aa}. The method is relatively easy to apply: using a set of pre-main sequence models, the code uses a Monte Carlo generator to draw an artificial population from an assumed initial Mass Function (IMF) with a total number of members input by the user, between a minimum and a maximum age, defined by model isochrones, properly shifted by the distance modulus of the complex\footnote{ The 0.107 kpc distance error quoted by Hachisuka (2006) for their maser parallax distances, translates into an error of about 0.2 mag in the brightness estimates. This is smaller than the size of our KLF bins (0.5 mag.) In terms of mass, this translates into an uncertainty of about 0.05 in a log-mass scale.}. For this study, we chose the isochrone set of \cite{DAntona:1997aa}, with a deuterium fraction $[D/H]=2\times 10^{-5}$. The user  has the option to choose the form of the IMF and also has the option to fix or to sample values for a set of parameters that includes an extinction distribution, a binary fraction, and the infrared excess fraction as a function of color ($H-K$). The peak and the slopes of the KLF are particularly sensitive to age and age spread, which are the two parameters that were left to vary. We decided to leave the binary fraction as a fixed, value of 20\%, based on estimations by \cite{Padgett:1997aa} suggesting that the binary fraction in the Orion B clusters may be about 15\% in separations up to 1000 AU --the separations we are able to resolve in our near-IR images. Also, \citet{Muench:2002aa} in a similar experiment showed that, particularly, the mean age and spread estimates were not significantly different when changing the binary fraction within 40 percent. A smaller fraction is probably too optimistic, but a fraction that is significantly large (30-40\%) tends to add dispersion in the histogram and ``dilutes" the location of the KLF peak. For the IMF, we decided to use the the parameters of the 3-part broken power law obtained for the IC 348 cluster by \citet{Muench:2003aa}\footnote{Notice that we cannot sample the whole mass spectrum at the distance of W3, specially given the high extinction. However, our $A_V$-limited samples are complete down to K=16.25 mag which at the distance of W3 (2.05 kpc) and given the estimated ages of the clusters, could range between 0.05 and 0.3 M$_\odot$, with an average around or below 0.1 M$_\odot$. This is below the expected peak of the IMF at 0.2 M$_\odot$ and that is why we consider necessary to use the three parts of the broken power-law parametrization. The interpolation code of \citet{Muench:2000aa} takes into consideration the minimum and maximum brightness of the cluster observations and samples the IMF within the proper ranges.}. Each of the fixed parameters will add its own error to the estimate of the best value for the running parameters, and there are also errors associated with the use of one or other isochrone model set. A full discussion of the method is out of the scope of this paper, however, the feasibility of this method has been properly discussed by \citet{Muench:2002aa} (chapter 5).

\begin{figure}
\centering
\includegraphics[width=3.25in]{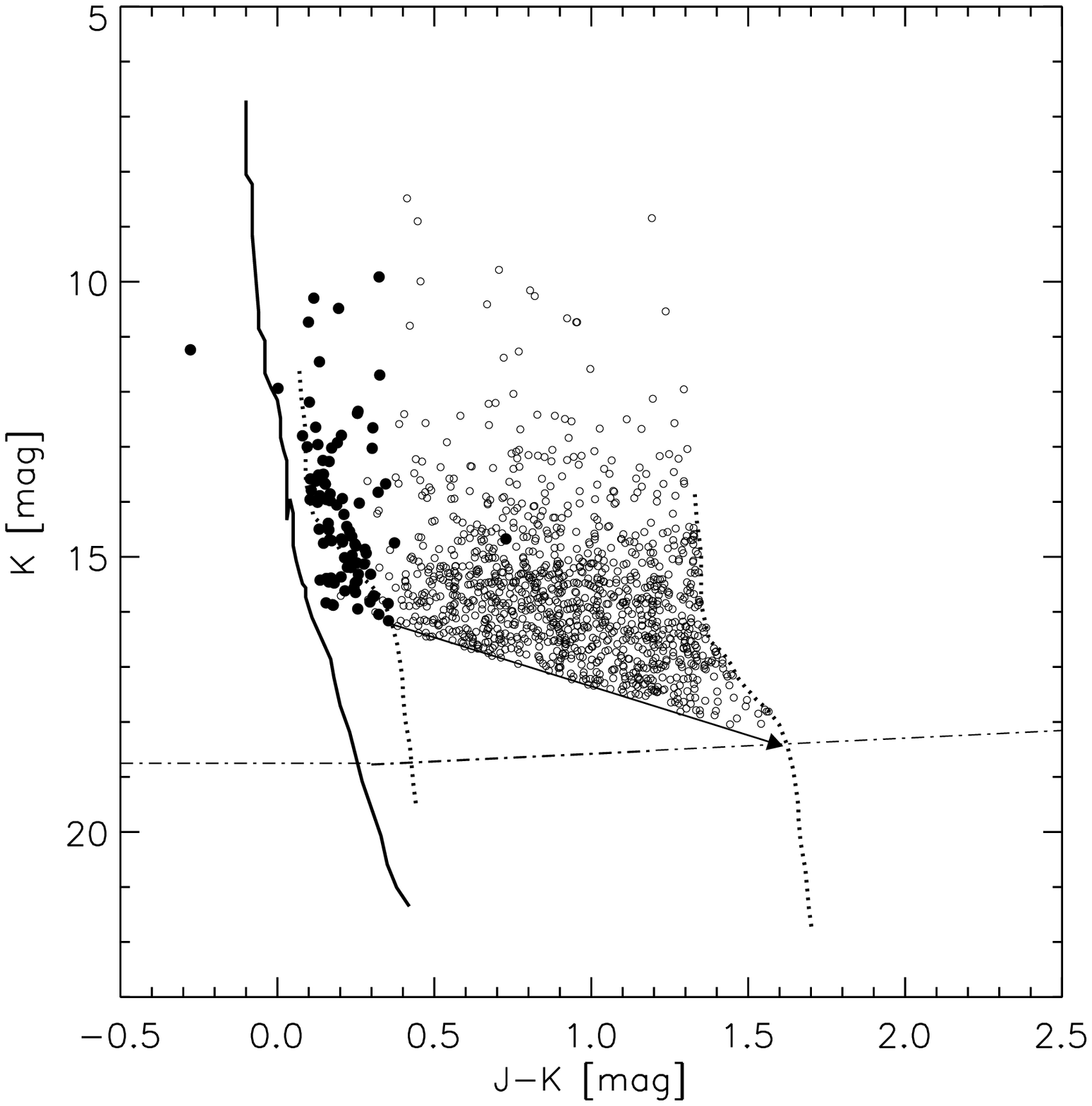}
\includegraphics[width=3.25in]{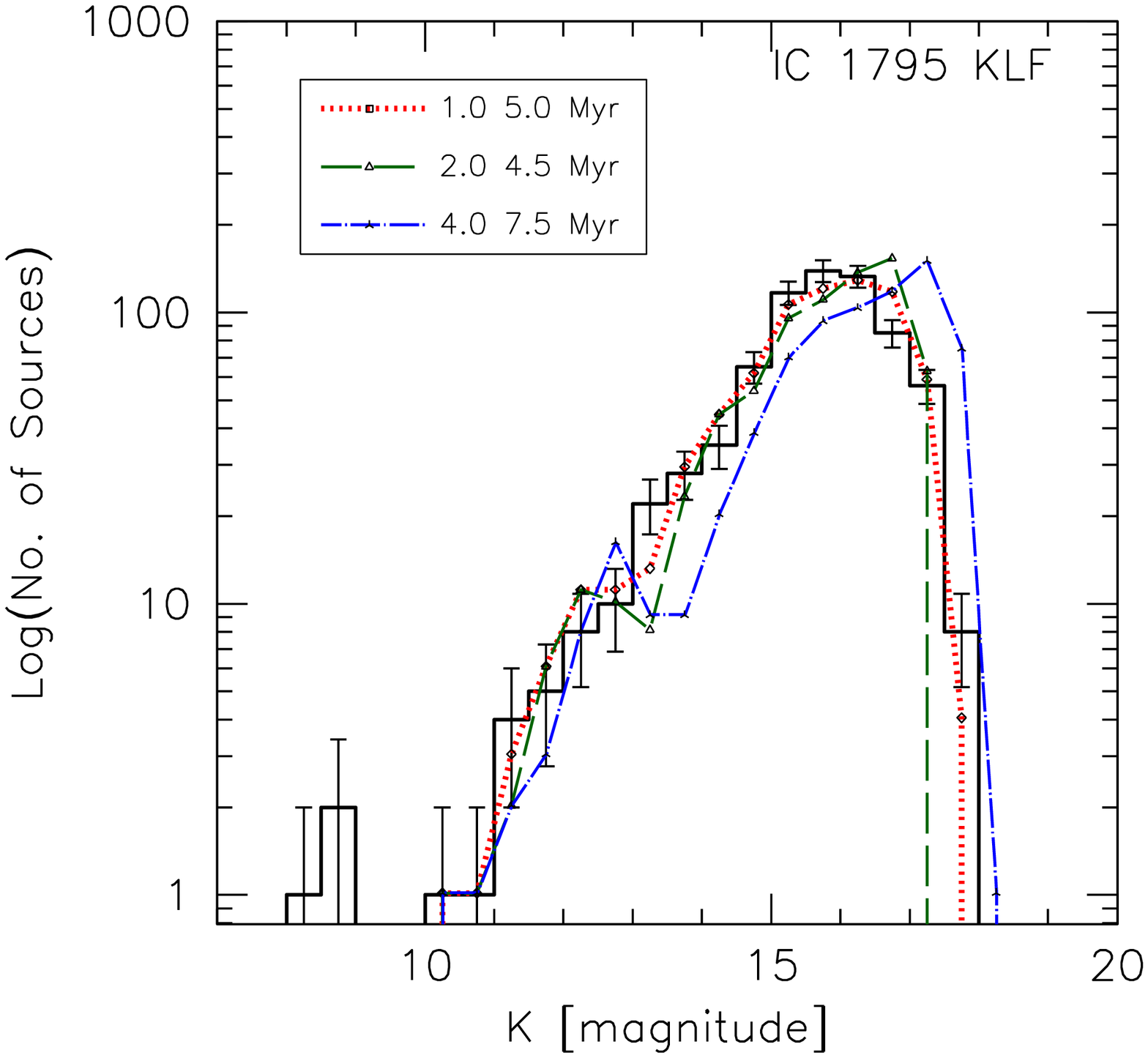}
\caption{
\textit{Left}: Extinction ($A_V$) limited selected sample for W3-Main using $K$ vs $H-K$ color-magnitude
diagram. The solid thick line is the zero-age main sequence. The two thick, dotted lines are a
3 Myr isochrone from the models of \cite{DAntona:1997aa}, before and after adding an extinction of
$A_V=20$ mag. This is also indicated by the extinction vector that runs from the lowest mass point
(0.3 M$_\odot$). The dot-dashed line indicates the sensitivity limit of our Calar Alto data set. Solid 
symbols indicate sources selected out as possible foreground sources. The open circle symbols indicate
sources selected as cluster members in the final sample. \textit{Right}: An illustration of KLF fitting, for IC 1795. The
solid, black histogram is the observed KLF, after subtraction of the field component; the red, green and blue
lines are artificial KLFs modeled after the properties of the cluster, using the code of \citeauthor{Muench:2000aa},
with reduced $\chi ^2$ probabilities of 0.95, 0.68 and 0.22, respectively.\label{fig:avlimitedsample}}
\end{figure}

\begin{figure}
\centering
\includegraphics[width=2.5in]{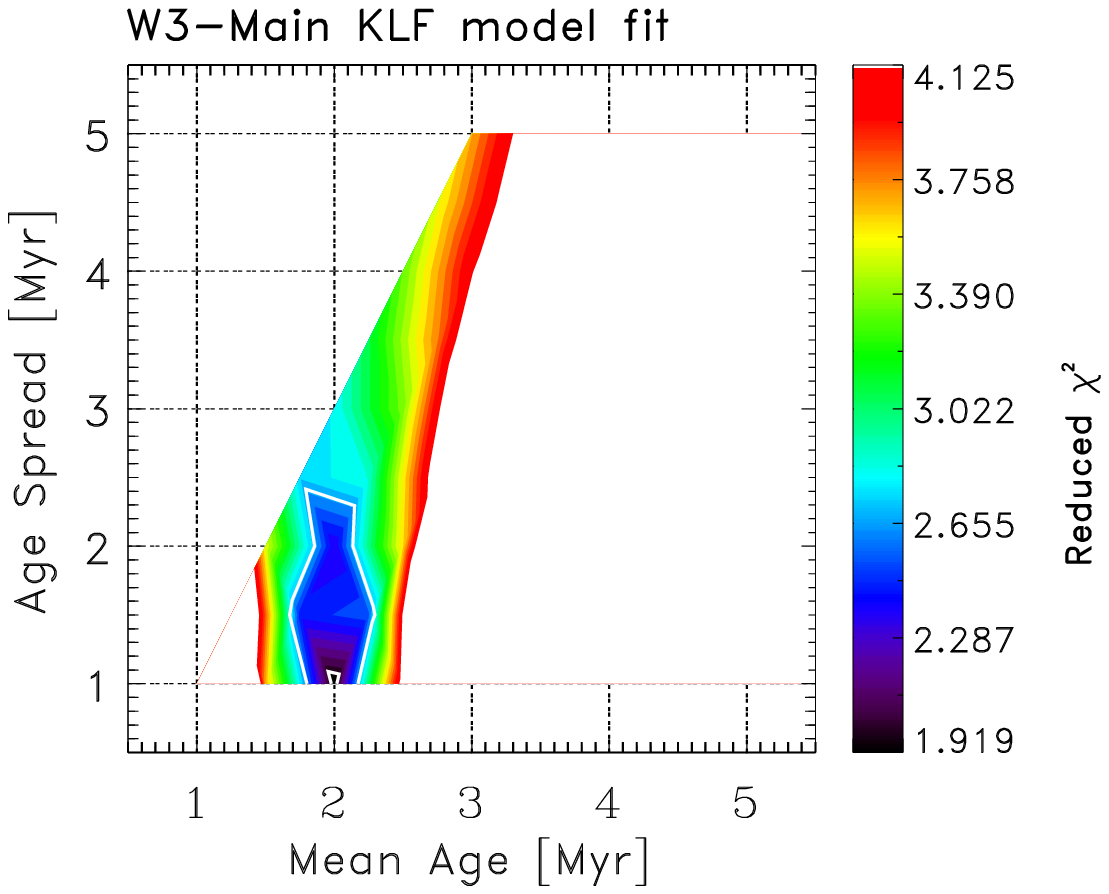}
\includegraphics[width=2.5in]{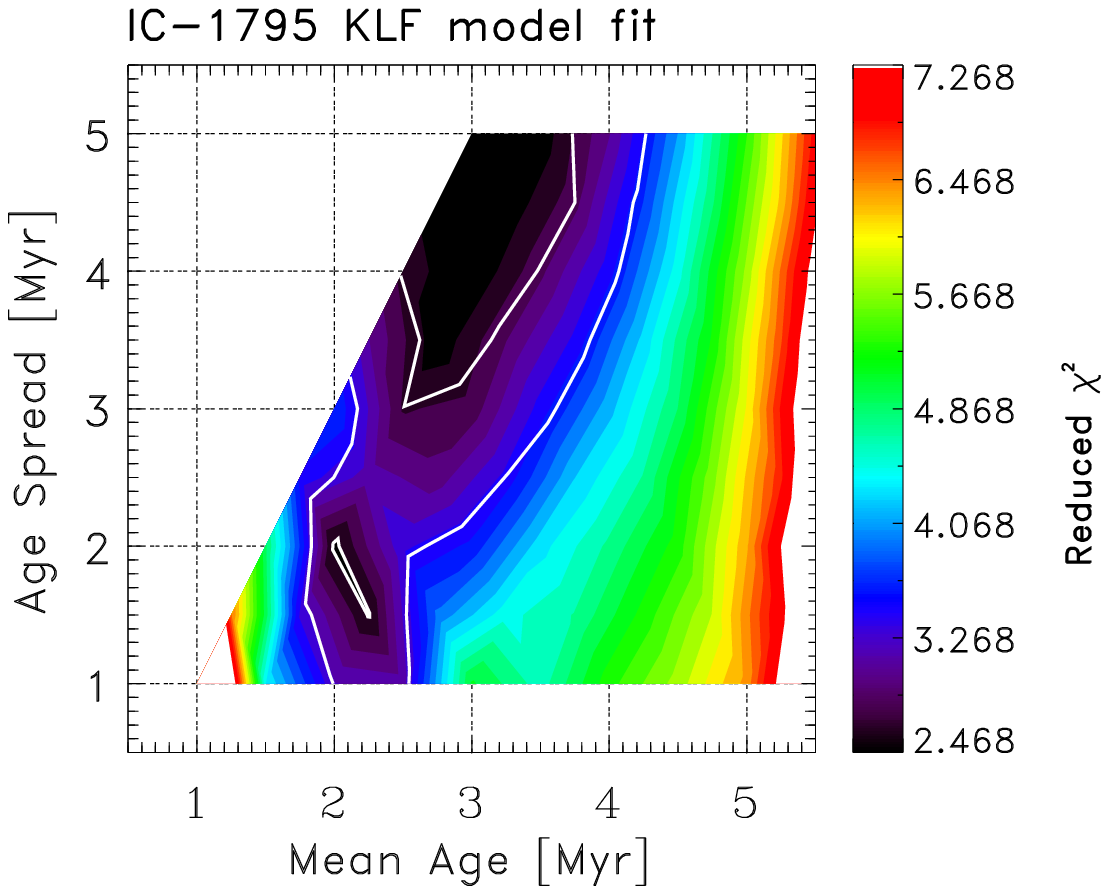}\\
\includegraphics[width=2.5in]{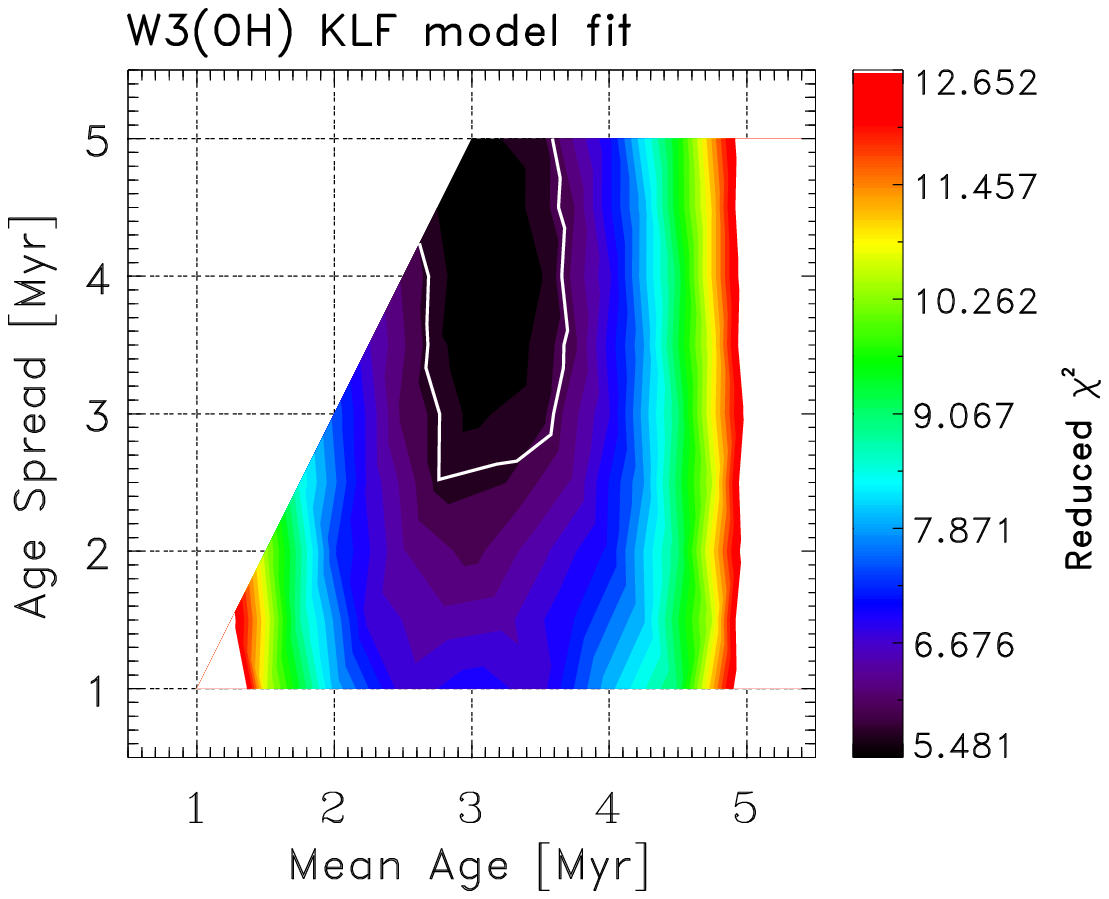}
\includegraphics[width=2.5in]{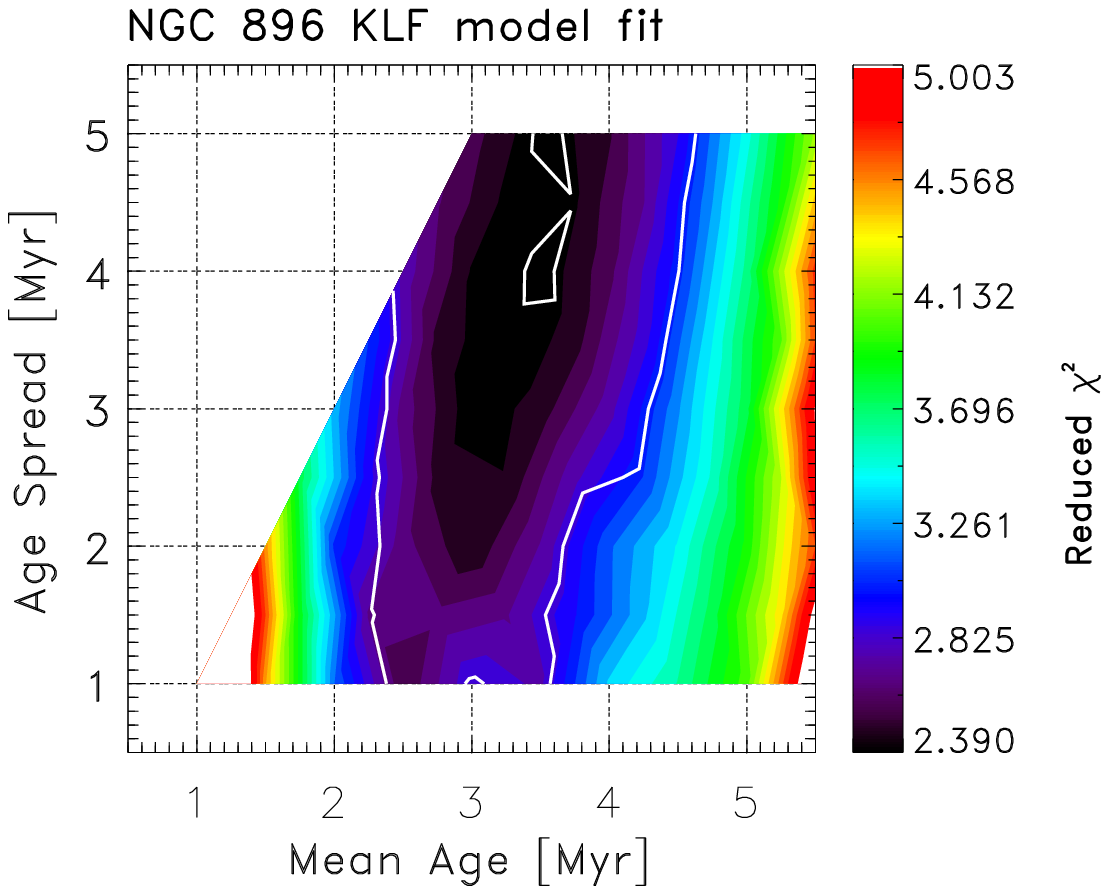}\\
\includegraphics[width=2.5in]{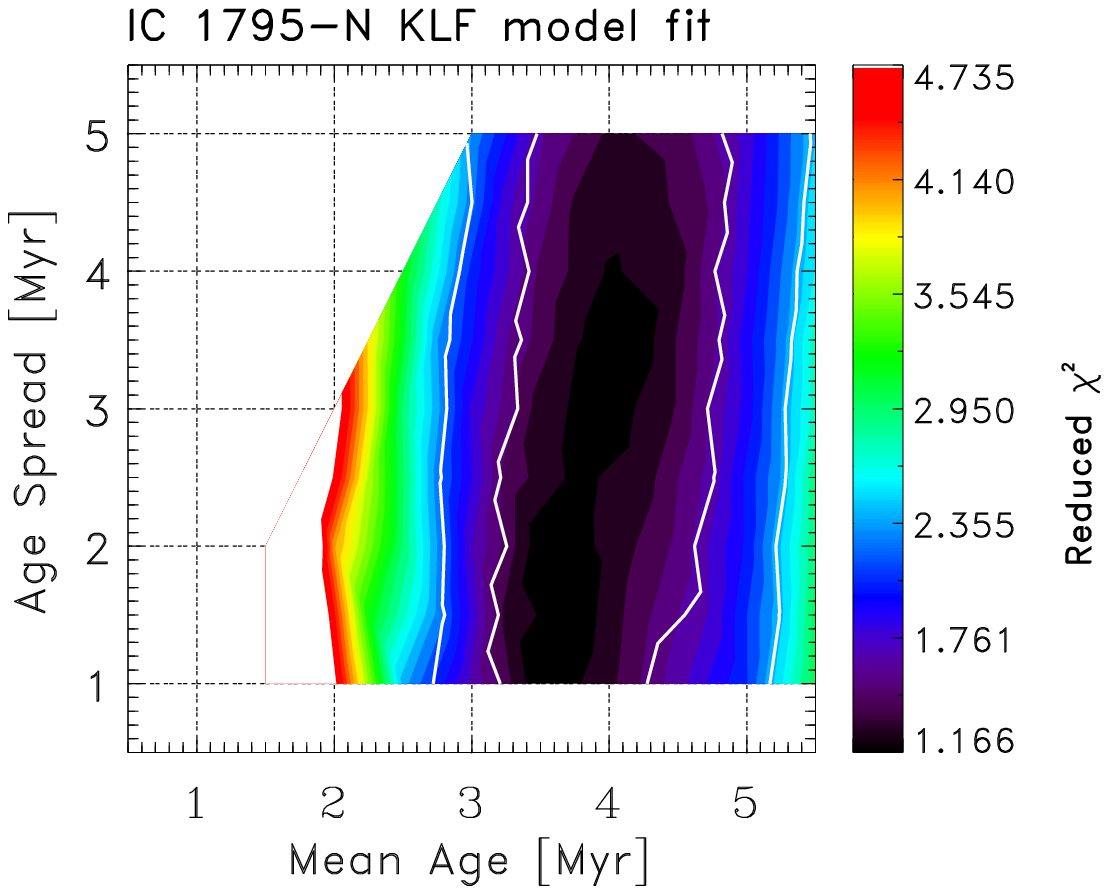}

\caption{ Age estimates for all principal clusters, illustrated
with a contour map of the normalized $\chi^2$ values. We plot the mean
age input for the artificial KLF models vs. the age spread used
in each simulation set. Areas at and below purple color levels indicate the
most likely values. The white contour lines trace the 68 and 95 percent confidence limits. 
The light line surrounding the contour areas indicates the limits of the model grid.\label{fig:agespread}}
\end{figure}

\par We made a grid of cluster models that varied the minimum age of the population between 0.5 and 10 Myr, in steps of 0.5 Myr, and having star formation age spreads of 1.0 to 5.0 Myr. For each minimum age and age spread value, we simulated 500 artificial KLFs and compared each of them with our observed KLF using a simple reduced $\chi^2$ scheme. Averaging over the simulations, we are able to determine which age or group of ages fits the observed KLF better (see Figure \ref{fig:avlimitedsample}). Our estimate of the age and age spread is defined by those models for which the confidence limit is above 68\% (2-sigma). In Figure \ref{fig:agespread} we show two examples of the age estimation scheme: we plot the mean age used in the simulation versus the width of the age spread. The reduced $\chi ^2$ value contours indicate the grid values for which the age models fit better (25 percentile). In these examples, the W3-Main cluster shows a very clear fit for a population with an age spread of less than 2.5 Myr, and a mean age of around 2.0  Myr. The IC 1795, instead, is less well constrained: we see two local $\chi ^2$  minima in the contour map. The first one agrees with a mean cluster age of less than 2.5 Myr old and an age spread of less than 3 Myr. The second minimum suggests a mean age between 2.5 and 3.5 Myr and an age spread between 3 and 5 Myr. The second option agrees much better with spectroscopic studies of IC 1795, particularly the one by \citet{Oey:2005ly}. To be conservative, we consider a mean age between 2.0 and 4.0 Myr, and an age spread in the whole range of 1.0 to 5.0 Myr. The main reason for the discrepancy in IC 1795 is the large number of very young (Class II) sources still present in the IC 1795 region, which is indicative of a mixed age population. The analysis suggests a very similar age and spread for W3(OH), which is surprising given that it hosts more embedded populations. The analysis for clusters IC 1795-N and NGC 896 yield mean ages from 2.5 to 4.5 and from 3.5 to 5.0 Myr with possible spreads from 2.0 to 5.0 and from 1.0 to 5.0 Myr, respectively. Unfortunately, the upper limits in the age spread are not well constrained for most clusters, however, age spreads much larger than 5 Myr sound little plausible for embedded star cluster populations. In table \ref{tab:principal_clusters} we list the main results of the artificial KLF modeling analysis for the five principal clusters. The age and age spread for each of the principal cluster is listed as the center and width of the top contour in the age spread vs age $\chi ^2$ maps.

\begin{center}
\begin{deluxetable}{llccccc}
\tablecolumns{8}
\tablewidth{0pc}
\tablecaption{Main Parameters and Age Estimates for Principal Clusters in W3\label{tab:principal_clusters}} 

\tablehead{
\colhead{Cluster} &
\multicolumn{2}{c}{Center Coords.} &
\colhead{$r_{cl}$ } &
\colhead{Disk Fraction\tablenotemark{b}} &
\colhead{Mean Age\tablenotemark{c}} &
\colhead{Age Spread\tablenotemark{c}}\\
\colhead{} &
\multicolumn{2}{c}{J2000} &
\colhead{[pc]\tablenotemark{a}} &
\colhead{[\%]} &
\colhead{[Myr]} &
\colhead{[Myr]} \\
}
\startdata

W3-Main    &  36.368671 & 62.097485 & 1.313 & 27$\pm$4 & 2.0 [1.5,2.5] & 1.5 [1.0,2.5] \\
IC 1795    &  36.652903 & 61.990908 & 1.244 & 29$\pm$5 & 3.0 [2.0,4.0] & 4.0 [1.0,5.0] \\
W3(OH)     &  36.798850 & 61.895349 & 1.644 & 27$\pm$3 & 3.0 [2.5,3.5] & 3.5 [3.0,5.0] \\
NGC 896    &  36.388061 & 61.996607 & 1.428 & 14$\pm$3 & 3.5 [2.5,4.5] & 4.0 [2.0,5.0] \\
IC 1795-N  &  36.569302 & 62.103874 & 1.618 & 12$\pm$2 & 4.0 [3.5,5.0] & 2.0 [1.0,5.0] \\

\enddata
\tablenotetext{a}{Assuming a distance $d=2.045$ kpc to the W3 Complex.}
\tablenotetext{b}{Estimated as the fraction of Class I+II sources
 within $R_{3\sigma}$, in the extinction-limited sample, down to $H_{dered}$=13.5 mag.}
\tablenotetext{c}{
The number pairs inside the brackets indicate the range of
model ages within the 2-sigma (68\%) confidence level. The number listed
before the bracket is the central value, which we list as the best estimate
for the age/age spread of the cluster}

\end{deluxetable}
\end{center}

\subsection{The Context of the W3 Molecular Cloud \label{s:analysis:ss:gas}}

\par The 1.1 mm Bolocam maps and the CO isotope radio emission maps of \citet{Bieging:2011kq} show that the IC 1795 is almost free of molecular gas emission, while most of the remaining cloud is concentrated on two large clump systems at W3-Main and W3(OH). The latter has a total mass of approximately $1.9\times 10^3\mathrm{\ M}_\odot$, while the former has a total mass of approximately $2.5\times 10^3\mathrm{\ M}_\odot$, divided in two main clumps named W3-East and W3-West \citep[][]{Rivera-Ingraham:2013fj}. In such layout, IC 1795 is seen surrounded by an almost symmetric shell of star forming molecular material. 

\par
\begin{figure*}
\centering
\includegraphics[width=7.0in]{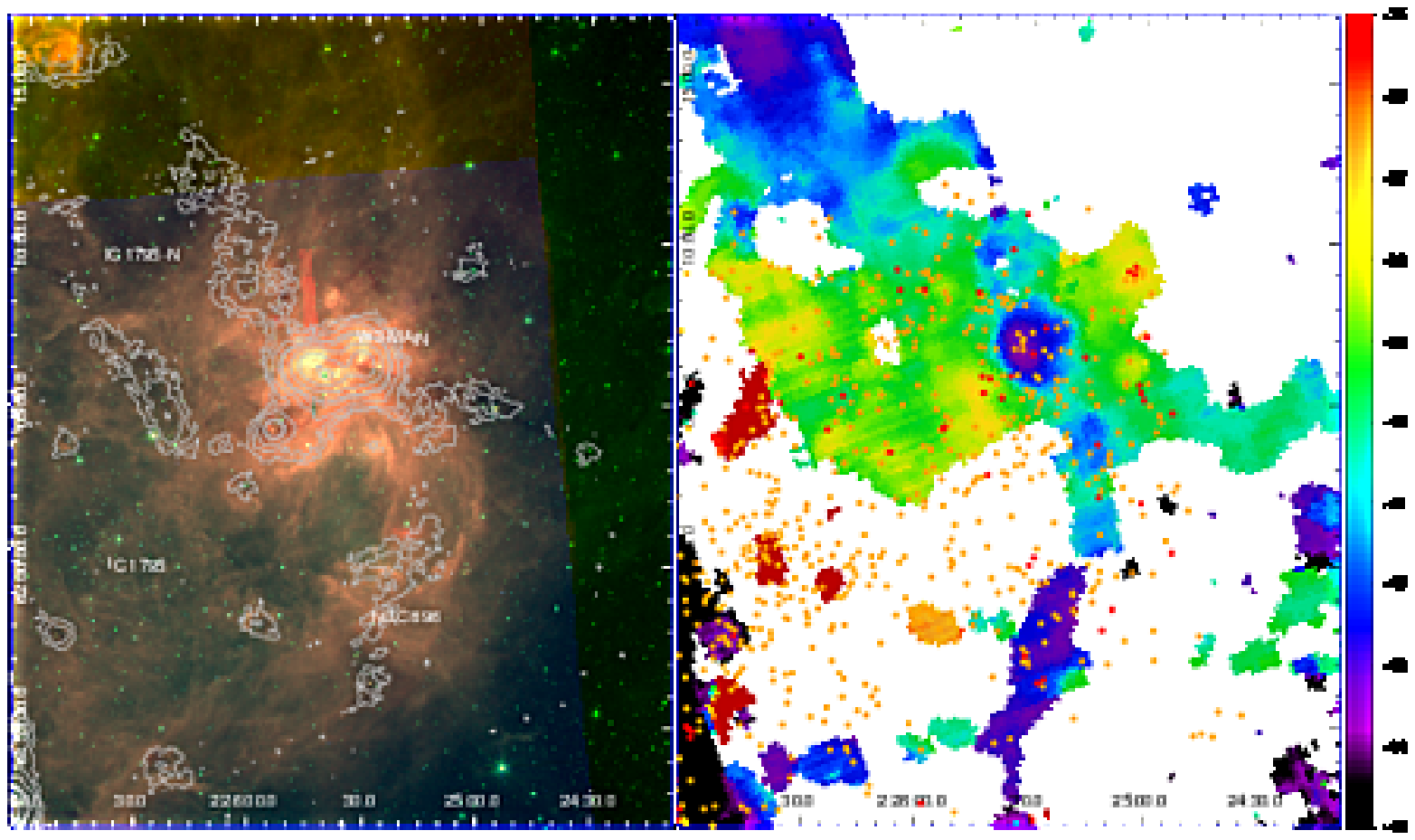}
\caption{Association of gas and stars in the W3-Main region. \textit{Left:} The RGB image in the background is 
constructed from Spitzer-IRAC [3.6],[5.0] and [8.0]$\mu$m
mosaics, and the contours indicate 1.1 mm emission in Jy/beam from the
Bolocam map at 0.12,0.24,0.48,0.96,1.2,2.4,4.8 and 9.6. \textit{Right:}
A first moment $^{13}$CO(2-1) map of the same region, from the dataset of \citet{Bieging:2011kq}. The color table indicates radial velocity in km/s. The red and orange dot symbols indicate the positions of Class I and Class II sources identified from our
main catalog, respectively. \label{fig:gas-stars-w3main}}
\end{figure*}

\begin{figure*}
\centering
\includegraphics[width=7.0in]{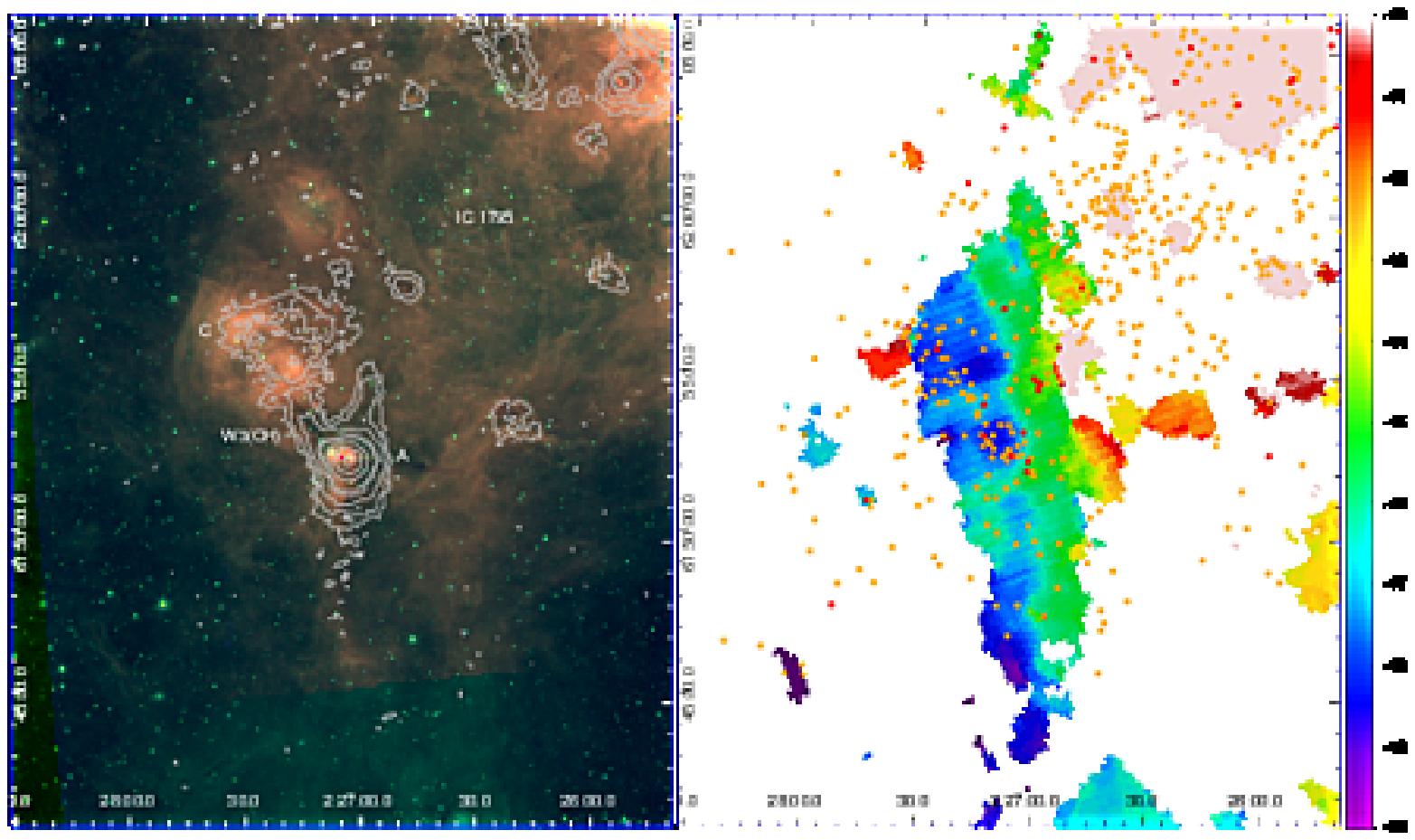}
\caption{Association of gas and stars in the W3(OH) region. Symbols and colors are equivalent as those in Fig. \ref{fig:gas-stars-w3main}. \label{fig:gas-stars-w3oh}}
\end{figure*}

 In Figures \ref{fig:gas-stars-w3main} and \ref{fig:gas-stars-w3oh} we show the locations of Class I and Class II source candidates in the context of the $^{13}$CO(1-0) first moment (average radial velocity) map of the cloud, as well as a map of 1.1 mm emission over infrared images at both the Eastern (W3(OH)) and Western (W3-Main) sides of the Cloud. 

\par The W3-Main map shows a very notorious blue-shifted peak at the center of the principal cluster, that coincides in velocity with a filamentary structure that extends south toward NGC 896 and north toward IC 1795-N. The blue-purple colored peak corresponds to a velocity range centered around -42.5 km/s, and it has been suggested to indicate central massive outflow activity at the center of the cluster \citep{Bieging:2011kq}. The main $^{13}$CO(2-1) emission structure associated with W3-Main with velocity range centered around -40 km/s (green colored in the image), shows a clumpy structure suggestive of other local velocity gradients. A few small receding clumps associated with velocities in excess of -36 km/s (red colored in the image) possibly represents gas still being expelled from the central cluster, and confirms that gas removal in IC 1795 was very recent. 

\par The W3(OH) map shows that the cluster formation activity at this side of the complex is associated with molecular emission with significant radial velocity gradients. This eastern side of the cloud has a filamentary morphology, with almost uniform radial velocity. There is a ~2 km/s difference between the east and the west edges of the filament, suggestive of a ``rolling" movement of gas in the north-south direction, but on the eastern side the peak velocity alternates several times from ~46 to ~48 km/s, indicating local gradients possibly linked to gas compression and expulsion at the W3(OH) groups. This is particularly evident at group W3(OH) A.

\par
\begin{figure}
\centering
\includegraphics[width=3.3in]{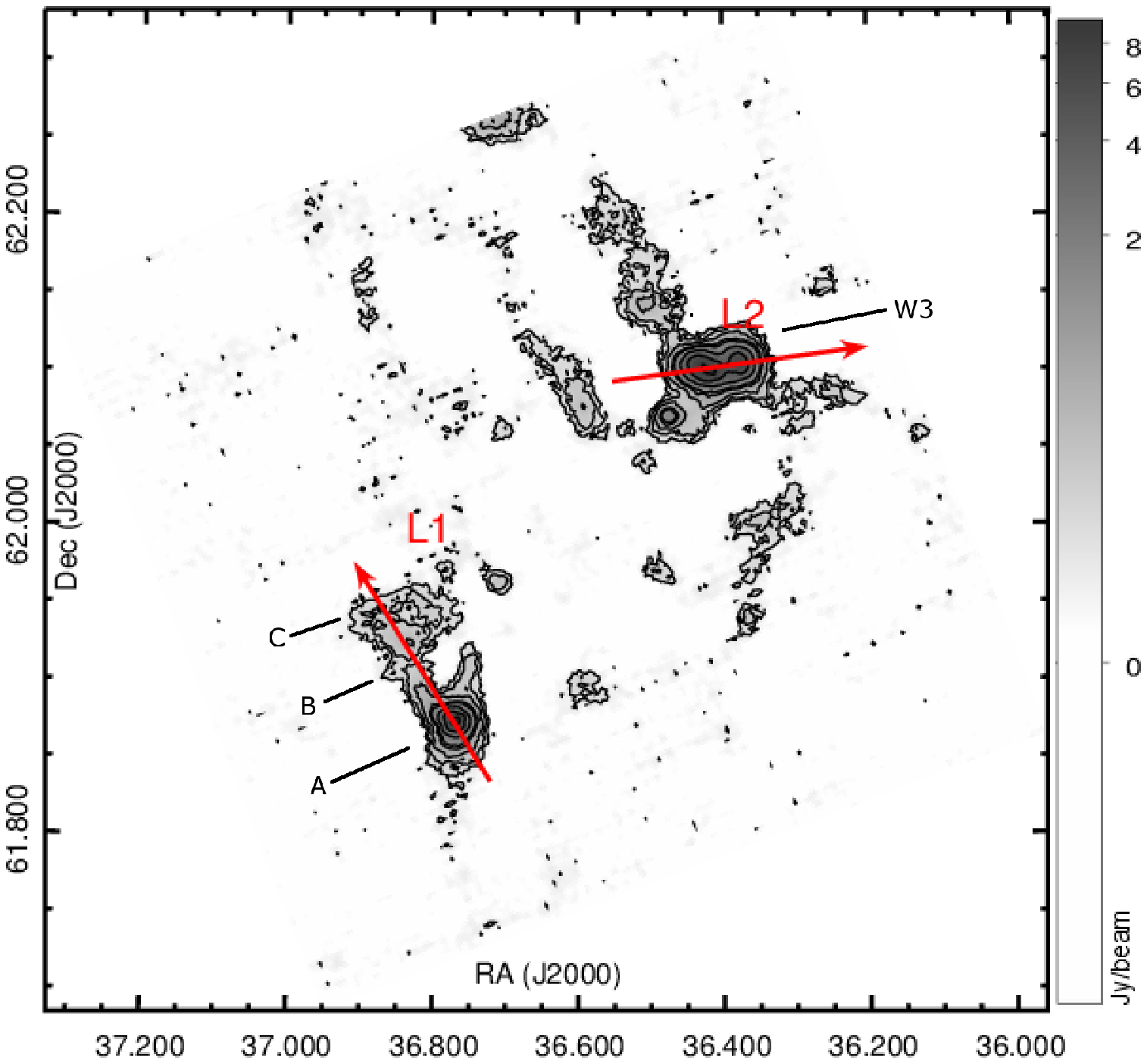}\\
\includegraphics[width=3.5in]{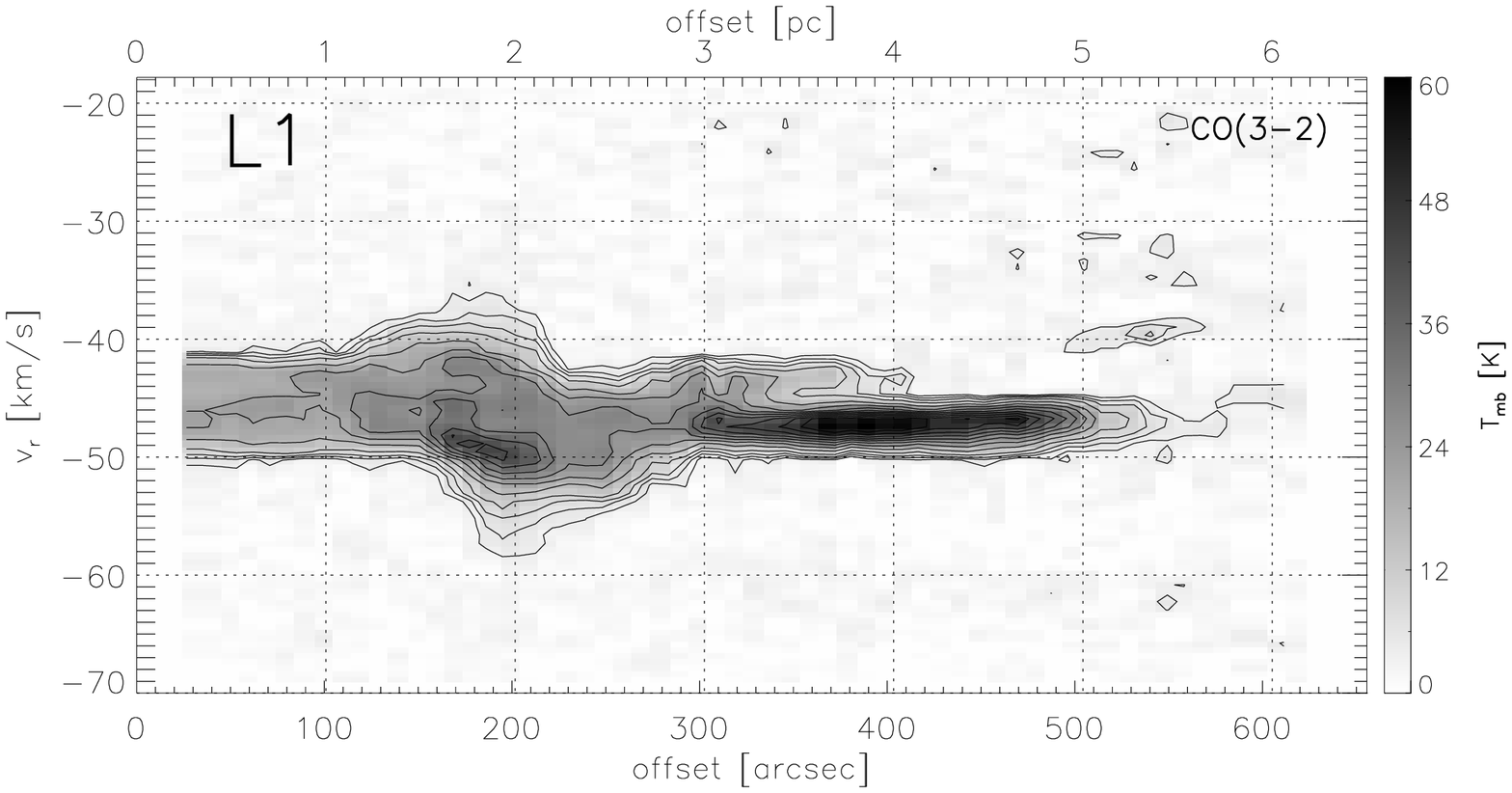}\\
\includegraphics[width=3.5in]{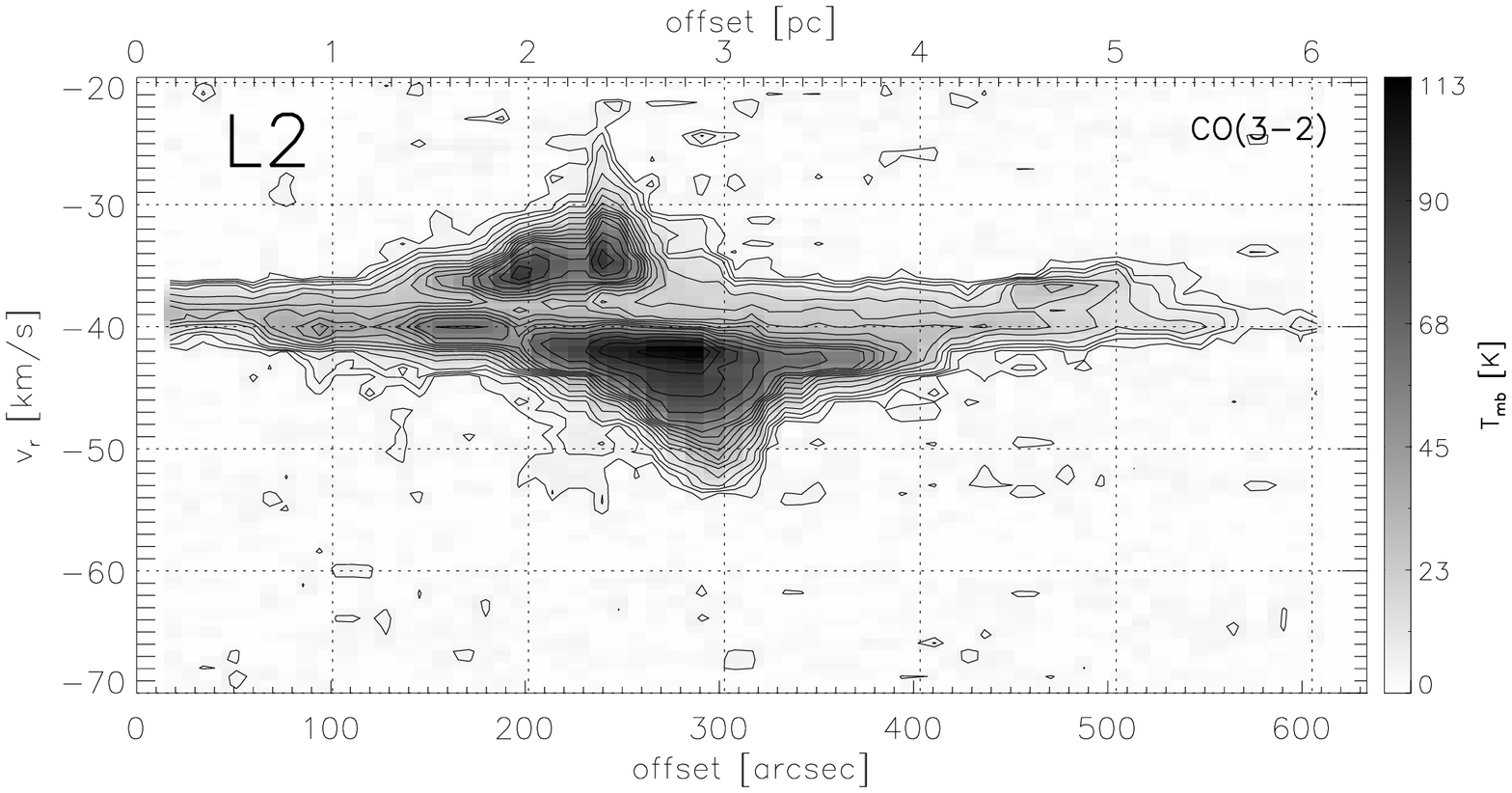}
\caption{Top: Bolocam 1.1 mm emission map (grayscale, overlaid with
same set of contours as in Fig. \ref{fig:gas-stars-w3main}). Arrows
indicate length and direction of two cuts used to create position-velocity
maps. Middle and Bottom: Position-velocity plots, from the
CO(3-2) emission maps of \cite{Bieging:2011kq}. The color
table indicates line temperature in K. \label{fig:velpos}}
\end{figure}

 In Figure \ref{fig:velpos} we show two position-velocity (p-v) plots corresponding to two cuts along the main structures of W3(OH) (cut $L1$) and W3-Main (cut $L2$) in the 1.1 mm map. We constructed these p-v plots from the $^{13}$CO(3-2) emission map of \cite{Bieging:2011kq}. 

\par The L1 plot shows a well defined bipolar structure at the location of cluster A with a peak to peak velocity difference of 15-20 km/s within less than 0.5 pc. The $^{13}$CO(3-2) map does not resolve the structure well enough to claim the presence of individual outflows, but the plot shows that the W3(OH) elongated clump may be associated with a structure coherent within (roughly) -50 and -45 km/s. The plot shows an opening flow of material associated with cluster A (offset$\approx 170\arcsec$), and a tighter blue-shifted flow associated with cluster B (offset$\approx 200\arcsec$). A third component may be present near offset$\approx 300\arcsec$, possibly associated with cluster C.

\par The L2 plot shows a very strong outflow feature associated with the main molecular gas clump in W3-Main, with a peak to peak velocity difference of 25-30 km/s within less than 0.7 pc. The red-shifted component is associated with the left lobe of the main molecular clump and vice versa. The two peaks of the red-shifted structure suggest a parabolic opening indicative of a widening outflow. The physical scales and velocity differences of these structures are very similar to those observed in other massive star forming regions like DR-21 \citep{schneider:2010aa}, G240.31+0.07 \citep{qiu:2009aa} or W49A \citep{peng:2010aa}. 

\par We estimated the integrated intensity of $^{13}$CO(3-2) in two square areas of $120\arcsec \times 120\arcsec$ at the center of the double peaked clump in W3-Main, which correspond to the blue and red peaks of the outflow feature. We defined these square areas from a first moment (mean velocity) map. Using the prescription of \citet{ginsburg:2011aa}, we estimated the column density, $N(H_2)$ and the total mass of outflow gas, $M_{out}$ enclosed in each the two areas (17.6 and 40.7 M$_\odot$ for the red and the blue peak, respectively). Then, following \citet{qiu:2009aa} we estimated the dynamical age, $t_{dyn}$ of each component (6.5 and 9$\times 10^4$ yr, for the red and blue peaks, respectively) and the mass loss rate, $\dot{M}_{out}=M_{out}/t_{dyn}$. We estimated that the total mass loss rate is close to $7.2\times10^{-4} \mathrm{M}_\odot \mathrm{\ yr}^{-1}$. This is close to estimates in other massive star forming regions, which have mass loss rates of $\sim 10^{-3} \mathrm{M}_\odot \mathrm{\ yr}^{-1}$. A mass loss rate like this, if constant, could remove $10^3\mathrm{\ M}_\odot$ of gas in about 1.5 Myr.

\section{Discussion \label{s:discussion}}

\par Our KLF analysis suggests that episodes of cluster formation in W3 may have started over 4 Myr ago, without any well defined pause. Actually, our data also suggests that star formation has been ubiquitous across the whole complex within the last 1.0-2.0 Myr:

\begin{enumerate}
\item All of the principal clusters exhibit a significant fraction of sources with detectable circumstellar emission. This is evidence in favor of recent formation across the whole complex. Class I sources are distributed preferentially in high extinction regions, indicating the most recent episodes of formation. However, Class III sources are present also near these highly embedded regions: this could be indicative of age spreads comparable to the T Tauri timescale (and possibly longer, if we consider young sources that escape classification via our color and X-ray emission scheme) across the whole complex. However, this could also be the result of rapid dissipation of disks, particularly for intermediate to high mass sources.

\item The age spread of IC 1795 cluster may be at least 4 Myr. This cluster also has a slightly larger density of Class III sources (e.g. 7 vs 5 percent of all YSO candidates down to a de-reddened brightness of H=13.5 mag, estimated within $R_{3\sigma}$ in IC 1795 and W3-Main, respectively). This is in agreement with previous studies that suggest this is the more evolved population in the complex. Also, the lack of nebulosity or strong CO emission is indicative of gas dispersal, which agrees with a more evolved cluster. However, the mean age that we estimate for this cluster is not very different from that of the youngest clusters in the complex, W3-Main and W3(OH). Also, the $R_{II:III}$ ratio in IC 1795 is comparable to W3-Main and W3(OH) and so is the total fraction of circumstellar disk sources in this cluster (about 30 percent within $r_{3\sigma}$[IC1795]). Still, no Class I sources are longer present in IC 1795. These results are not contradictory. They all suggest that the forming gas dispersal timescale in IC1795 was rapid, comparable or faster than the T Tauri timescale.

\item The KLF analysis suggest that W3(OH) has an age and age spread not very different from IC 1795. Again, this is indicative of more than one episode of formation. The main groups A, B and C in W3(OH) are still embedded in bright nebulosity and the principal cluster has a large circumstellar disk fraction, comparable to IC 1795 and W3-Main. There is also a significant number of Class I sources in W3(OH). However, there may be slightly older groups toward the eastern edge of the complex, judging from the presence of Class II and Class III sources to the edge of the molecular clump and no nebulosity. The $R_{II:III}$ maps show an abrupt termination at the eastern side of the W3(OH) principal cluster which reinforces this idea. The structure in W3(OH) (specially if there is an older component) weakens the argument of a Gaussian cluster used for the GMM analysys. We need deeper photometric data in order to reconcile this problem.

\item The KLF modeling analysis shows that IC 1795-N and NGC 896 may host the oldest clusters in the complex with a mean age between 3.5 and 4 Myr. These two aggregations could have been formed in an external layer of the original molecular cloud, flanking IC 1795 from north and south.

\item Our analysis suggests that W3-Main hosts the most recent cluster formation episode, starting just over 2.0 Myr ago and having the shortest spread, with less than 2.5 Myr. 
\end{enumerate}

\par
\begin{figure*}
\centering
\includegraphics[width=6.0in]{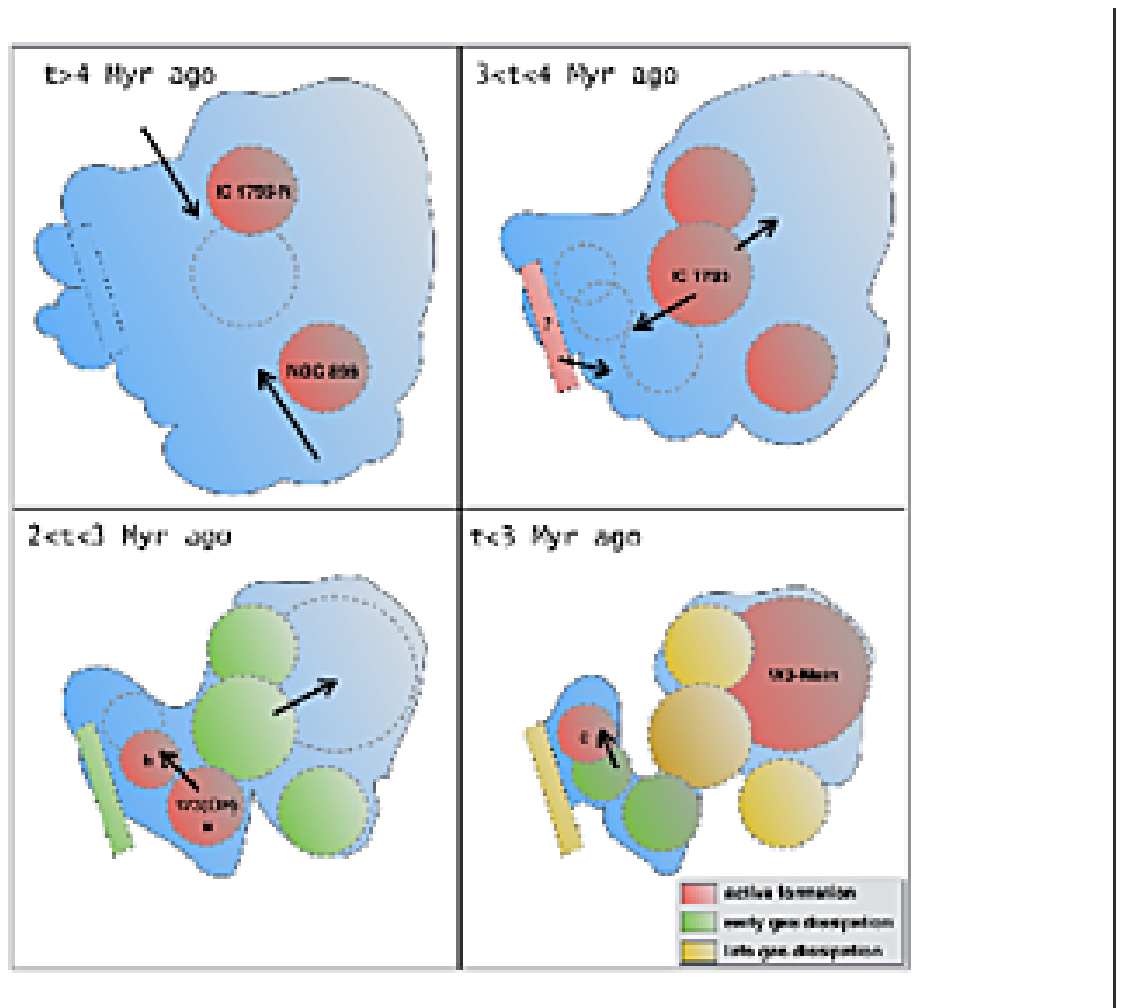}
\caption{A cartoon illustrating our proposed scheme for the
history of star formation in the W3 Complex. The blue regions
indicate the extension of the molecular gas cloud. Dotted, empty
regions indicate regions of collapse towards cluster formation. Red
shaded areas indicate active cluster formation regions. Green shaded areas
indicate cluster regions where gas dissipation has started and yellow shaded
areas indicate slightly older clusters where gas dispersal has removed a significant
amount of the parental gas and the cluster. The rectangular, easternmost area,
represents a region of earlier formation with unknown extension, that might contribute today to a larger age spread observed in the W3(OH) cluster complex. Black arrows indicate the suggested progression of cluster formation in the complex.
 \label{fig:cartoon}}
\end{figure*}

 In Figure \ref{fig:cartoon} we present a cartoon illustrating one possible scenario of the progression of cluster formation in W3 based on our KLF analysis. We suppose that the original molecular cloud might have been an area somewhat larger than the current span of the cluster complex, over 4 Myr ago. As shown by \citet{Thronson:1985aa}, the W3 complex formed in a ridge of swept up molecular gas in the western age of the giant HII region W4. The expansion of W4 may have triggered the collapse of the W3 molecular ridge. By then, the formation of IC 1795-N and NGC 896 might have started, while IC 1795 and an area of unknown extension near the eastern edge of W3(OH) (we indicate this region symbolically in the cartoon) started to collapse and form stars. Just over 3 Myr ago, the process of cluster formation continued with IC 1795 and the first episode in W3-Main and W3(OH), while clumps corresponding to the present W3(OH) cluster group would be collapsing. About 2-3 Myr ago, gas dispersal could have started in IC 1795, IC 1795-N, NGC 896 and the eastern edge of W3(OH). The most recent episodes in W3-Main and W3(OH) would have started. By then, gas would be dispersed away from the center of the complex and start concentrating at the W3(OH) and W3-Main regions. This would resemble the current layout of the complex, with most of the remnant molecular gas being at W3(OH) and W3-Main. In this scenario, cluster formation progressed first from north and south towards the center of the complex and then from the center towards east and west.It is important to notice that this scenario is based on our estimates of the mean ages of the clusters, but could only be supported by a better constrain of the age spreads of the clusters. 
 
\par One important point to discuss here is whether or not induced formation plays a dominant role on the formation of clusters in the W3 complex, as suggested by previous studies. For instance, \cite{Ruch:2007fk} support the picture in which IC 1795 triggered the formation of W3 Main and W3(OH) episodes. The study by \cite{Rivera-Ingraham:2011yq} suggest that triggering processes were important in the formation of the clusters flanking IC 1795, but also that triggering processes work at local (sub-parsec) scales, with high mass stars acting to confine and compress material, enhancing the efficiency for the formation of new high mass stars by making convergent flows. This, in turn, would help to the progression of formation from the outer toward in the inner regions of the complex. The study by \cite{Feigelson:2008vn}, pointed to significant gradient of ages between the central OB sources of the cluster and a widely distributed population of PMS stars.

\par What does our study tell about the global star forming history of the complex? On one hand, if gas removal in the central cluster IC 1795 contributed to collect gas in a shell like structure, it may have enhanced the efficiency of formation in W3-Main and W3(OH). This would be in agreement with the flow convergence scenario discussed by \citeauthor{Rivera-Ingraham:2013fj}. According to that picture, IC 1795 could have been the major source of induction for star formation in the molecular shell.

\par Our KLF model analysis indicates large age spreads across the complex, with a progression that lead to cluster formation across the entire complex in a period of about 3 million years. This is in good agreement with our analysis of the YSO class surface density ratio across the complex. In that case, the progression is possibly related to primordial density structure of the cloud. This is in agreement with the scenario proposed for other complexes like the Rosette Molecular Cloud \citep{Roman-Zuniga:2008aa,Wang:2009aa,Ybarra:2013kh}, where it has been shown that progression of cluster formation could rather follow the density structure of the cloud than a process of induction in a sequence from the older to the younger episodes. 

\par Our YSO surface density analysis provides evidence of sub-structure in the principal cluster populations. Small aggregations of stars around the main cluster events (very evident in W3(OH), and groups like R102 and R103), revealed as significant peaks in the surface density of YSOs, in many cases coincident with the locations of Class 0/I candidates. This is suggestive of a complex scenario, in which stellar collections of different sizes may form with small age differences by the compression of gas in various locations within large clumps. Strong evidence of sub-clustering in W3-Main was also provided by the very deep near-IR KLF by \cite{Bik:2014aa}: they pointed out that an age gradient was evident from edge to center in W3-Main, judging from a higher fraction of circumstellar disks near the center. The large number of disk bearing sources in IC 1795 could be indicative of a similar process, where cluster formation occurs in sub-structures. If these sub-structures do not form simultaneously, they certainly may lead to a large age spread, as observed. 

\par The projected separations between surface density peaks in the \texttt{gather} map range is 1.4$\pm$0.4 pc. These separations are comparable to the estimated radii of the principal clusters identified with our GMM model. This scale is shorter than the Jeans length (2.0 pc) for a mean density of $10^2 \mathrm{cm}^{−3}$ and $T_K = 10$ K, typical of the diffuse gas traced by $^{12}$CO and may be tracing the primordial fragmentation of the cloud. One possibility is that density peaks in the \texttt{gather} map are sub-clusters that eventually merge into larger groups, but that will depend on the dynamical stability of the aggregations and the gravitational potential of the corresponding principal cluster. Numerical simulations \citep{Bonnell:2011kx,Allison:2010ca} suggest, that clusters are initially highly structured, but rapidly evolve by merging into large entities. On the other hand, small groups of stars may disperse rapidly or even remain stable depending crucially on the number of members and the star forming efficiency \citep{Lada:2003aa,Adams:2001aa}.  It is also expected that most small, embedded aggregations end up being disrupted early, as proposed by the ``infant mortality" and ``cruel cradle" models. 

\par The formation of small groups that disperse rapidly and merge into a larger central structure could provide a simple explanation for the presence of larger age spreads in a lower surface density population in regions like IC 1795-N or NGC 896, and would be in agreement with age spreads and sub-structure in the younger principal clusters.

\par The strong outflows in the molecular gas shell at W3-Main and W3(OH) confirm a very rapid removal of gas in the complex. This removal may occur in a period comparable to the T Tauri timescale, as evidenced by Class II source candidates in IC 1795, where gas is mostly dispersed. However, it is important to stress that the mass loss rate cannot be constant, and will probably decline within a few $10^5$ Myr, if it has not already done so. The estimate we provided, however, may not be far from the average mass loss rate during the main gas dissipation period.

\section{Conclusions \label{s:conclusions}}

\par We obtained deep, high resolution ground-based near-IR images of the Eastern edge of the W3 region in the Perseus Arm. These data were combined with archive imaging data from the Spitzer and Chandra space telescopes to obtain a new catalog of young sources for the W3 Complex. We used our resultant catalog to: 

\begin{enumerate}
\item Identify and classify young (Class I, II, III) embedded sources in the region.
\item Determine the locations and extension of the principal clusters in the complex. 
\item Determine the spatial distribution of young sources across the complex. This distribution shows significant substructure.
\item Construct extinction limited samples that allowed us to obtain clean K-band luminosity functions of the principal clusters. We compared the resultant functions with those from artificial clusters of ages created from pre-main sequence models. This analysis permitted to estimate mean ages and possible age spreads for the principal clusters of the W3 Complex. 
\end{enumerate}

\par We found that IC 1795, the central cluster of the complex, still has a large fraction of Class II sources, despite not being embedded anymore, like clusters W3-Main and W3(OH). The $R_{II:III}$ ratio is large across IC 1795, just as it is in the flanking embedded clusters. The $^{13}$CO map shows some receding molecular gas clumps coincident with the center of IC 1795. This, along with the absence of Class I sources suggest that parental gas was removed quickly from IC 1795, in a period shorter than the T Tauri timescale. The high mass loss rate estimate at the center of W3-Main supports this strongly.

\par Our Gaussian Mixture Model analysis clearly identified the well known clusters IC 1795, W3-Main and W3(OH). In addition, two principal clusters could be identified, one associated with the compact HII region NGC 896 and one more located north of IC 1795. 

\par We estimated the ages and age spreads of the principal clusters by comparing their K-band luminosity functions with those of artificial young clusters. The differences between the ages of the clusters are not significantly large. We also were able to constrain the age spread of the W3 cluster, but we could not constrain well the age spread of the other four clusters, so there might be overlaps in the formation periods of all clusters in the complex. This, is confirmed by a large number of Class II and Class III sources across the whole complex. This suggest a progression of cluster formation  scenario in which formation proceeded as a result of primordial density structure of the cloud. 

\par Using our list of Class II candidates, we constructed a surface density map that shows a significant amount of substructure in the principal clusters of the complex. The large age spread in the clusters may indicate that formation proceeds in episodes. If  sub-structures observed do not form simultaneously, they certainly may lead to a large age spread, like we observe. 

\par We propose a scenario in which star cluster formation progressed over 4 Myr ago from north and south (clusters NGC 896 and IC 1795-N) toward IC 1795. Later (about 2 to 3 Myr ago), the progression moved from center (IC 1795) towards East and West (W3(OH) and W3-Main). We also suggest that an undefined region East of W3(OH) could form groups of stars before the formation of the embedded multiple embedded groups we see today. 

\par The formation of W3(OH) and W3 Main is still in progress at the edge of the giant molecular shell of the complex. Using molecular emission maps, we showed that the two main gas clumps associated with these clusters (which together contain more than 4.5$\times 10^3\mathrm{\ M}_\odot$) show strong velocity gradients and gas outflow activity, spatially coincident with clusters and sub-structures. This is in good agreement with the convergent flow scenario \citep{Rivera-Ingraham:2013fj} that contributes to a large massive star formation efficiency in W3. Our analysis suggest that both the progression of formation following the primordial structure of the original Giant Molecular Cloud, and the triggered formation caused by the compression and dispersal of gas (that is, within the cloud) are both important in W3.

\par The W3 Complex, presents a rather complicated layout. The principal clusters  overlap partially along the line of sight and they may still interact with each other. Moreover, the clusters present a significant sub-structure. If the sub-structures are bonafide sub-clusters, they pose an interesting problem for future higher quality data. Particularly, space based  mid-infrared imaging with high resolution and deeper photometric limits (e.g. the James Webb Space Telescope) will provide more complete samples of the sub-cluster population. Also, spectroscopic data for individual young candidate members, may refine age and age spread estimates. 

\acknowledgments

\par \textbf{Acknowledgements:} We thank an anonymous referee for comprehensive and useful comments that greatly improved the content of our manuscript. CRZ acknowledges support from CONACYT project CB2010-152160, Mexico and program UNAM-DGAPA-PAPIIT IN103014. JY and MT acknowledge support from program UNAM-DGAPA-PAPIIT IN101813. GMV acknowledges support from an introduction to research scholarship from the Programa de Introducci\'on a la Investigaci\'on de la Junta de Ampliaci\'on de Estudios (JAE INTRO) CSIC/Instituto de Astrof\'isica de Andaluc\'ia. EAL acknowledges support from the National Science Foundation through NSF Grant AST-1109679 to the University of Florida.

\par This study is based on observations collected at the Centro Astron\'omico Hispano Alem\'an (CAHA) at Calar Alto, operated jointly by the Max-Planck Institut f\"ur Astronomie and the Instituto de Astrof\'isica de Andaluc\'ia (CSIC). We acknowledge the staff at Calar Alto for top of the line queued observations at the 3.5m with OMEGA 2000. We acknowledge use of data products from the 2MASS, which is a joint project of the University of Massachusetts and the Infrared Processing and Analysis Centre/California Institute of Technology (funded by the USA National Aeronautics and Space Administration and National Science Foundation). This work is partly based on observations made with the Spitzer Space Telescope, which is operated by the Jet Propulsion Laboratory, California Institute of Technology under a contract with NASA. The scientific results reported in this article are based to a significant degree on data obtained from the Chandra Data Archive; particularly, we made use of data obtained from the Chandra Source Catalog, provided by the Chandra X-ray Center (CXC) as part of the Chandra Data Archive. In this paper we discuss observations from the Bolocam Galactic Plane Survey (BGPS; PI John Bally); the BGPS project is supported by the National Science Foundation through NSF grant AST-0708403. In the present paper we discussed observations performed with the ESA Herschel Space Observatory \citep{Pilbratt:2010en}, in particular employing Herschel's large telescope and powerful science payload to do photometry using the PACS \cite{Poglitsch:2010bm} and SPIRE \cite{Griffin:2010hz} instruments. Herschel is an ESA space observatory with science instruments provided by European-led Principal Investigator consortia and with important participation from NASA. This research made use of Montage, funded by the National Aeronautics and Space Administration's Earth Science Technology Office, Computation Technologies Project, under Cooperative Agreement Number NCC5-626 between NASA and the California Institute of Technology. Montage is maintained by the NASA/IPAC Infrared Science Archive. We made use of the \texttt{pvextractor} tool by Adam Ginsburg, that is part of the Radio Astro Tools repository (http://github.com/radio-astro-tools).



{\it Facilities:} \facility{CAHA:3.5m (OMEGA 2000)}, \facility{Spitzer}, \facility{CXO}, \facility{CSO}, \facility{Herschel},  \facility{ARO:12m}






\appendix

\section{Near-IR Observations \label{app:obs}}

\par In Table \ref{tab:obs} we list information corresponding to our near-IR
Calar Alto 3.5m OMEGA 2000 observations.


\begin{deluxetable}{llcccccl}
\tablecolumns{8}
\tablewidth{0pc}
\tablecaption{Near-Infrared Observations of W3 Fields\label{tab:obs}} 
\tablehead{
\colhead{Field ID} &
\colhead{Date Obs.} &
\multicolumn{2}{c}{Center Coords.} &
\colhead{Filter} &
\colhead{Seeing} &
\colhead{BD Peak\tablenotemark{a}} \\
\colhead{} &
\colhead{} &
\multicolumn{2}{c}{J2000} &
\colhead{} &
\colhead{[$(\arcsec)$]} &
\colhead{[mag]} \\
}
\startdata

\multicolumn{7}{c}{OMEGA 2000-CAHA 3.5m OBSERVATIONS}\\*
\cline{1-7}\\*

CAHA-W301   &   2008-11-08   &  36.580583 & 62.135000  &    $J$   &	1.00  &  21.00 \\
CAHA-W301   &   2008-11-08   &  36.580625 & 62.135000  &    $H$   &	1.05  &  20.00 \\
CAHA-W301   &   2008-11-07   &  36.580625 & 62.135000  &    $K$   &	1.07  &  19.50 \\

CAHA-W302   &   2008-11-09   &  37.100042 & 62.108056  &    $J$   &	1.09  &  20.50 \\
CAHA-W302   &   2008-11-08   &  37.100042 & 62.108056  &    $H$   &	0.98  &  20.25 \\
CAHA-W302   &   2008-11-09   &  37.100042 & 62.108056  &    $K$   &	1.00  &  19.25 \\

CAHA-W303   &   2008-11-09   &  36.659000 & 61.985833  &    $J$   &	1.08  &  20.75 \\
CAHA-W303   &   2008-11-09   &  36.659000 & 61.985833  &    $H$   &	1.07  &  19.75 \\
CAHA-W303   &   2008-11-09   &  36.658958 & 61.985833  &    $K$   &	1.16  &  19.25 \\

CAHA-W304   &   2008-11-09   &  37.056875 & 61.904167  &    $J$   &	1.04  &  20.75 \\
CAHA-W304   &   2008-11-09   &  36.580583 & 62.135000  &    $H$   &	0.96  &  20.50 \\
CAHA-W304   &   2008-11-09   &  37.056875 & 61.904167  &    $K$   &	0.91  &  19.50 \\

CAHA-W3CF\tablenotemark{b}   &   2008-11-09   &  33.924208 & 60.545000  &    $J$   &	1.07  &  21.00 \\
CAHA-W3CF   &   2008-11-09   &  33.924208 & 60.545000  &    $H$   &	1.14  &  19.75 \\
CAHA-W3CF   &   2008-11-09   &  33.924250 & 60.545000  &    $K$   &	1.06  &  19.50 \\

\enddata
\tablenotetext{a}{Brightness Distribution Peak. Expresses turnover point of observed magnitude distribution} 
\tablenotetext{b}{Control Field}
\end{deluxetable}

\end{document}